\documentclass[aps,prb,twocolumn,nobalancelastpage,%
footinbib,%
noshowpacs,preprintnumbers,superscriptaddress%
]{revtex4-1}
\usepackage{amsmath,amssymb,amsfonts,fixmath,siunitx}
\usepackage{bm,graphicx,color}
\usepackage[pdftex,colorlinks=true,
					 bookmarksnumbered,bookmarksopen=true,bookmarksopenlevel=-1]{hyperref}

\def\dag{^{\dagger}}

\newcommand{\bra}[1]{\langle #1 \vert} 
\newcommand{\ket}[1]{\vert #1 \rangle} 
\newcommand{\ip}[2]{\langle #1 \vert #2 \rangle}  

\def\c{\gamma}
\def\d{\delta}

\def\e{\epsilon}
\def\ve{\varepsilon}

\def\k{\kappa}

\def\s{\sigma}

\def\w{\omega}

\newcommand{\abs}[1]{\left\vert #1\right\vert}
\newcommand{\appref}[1]{Appendix \ref{#1}}

\def\cp{\citep}

\def\ct{\citet}

\def\eg{e.g.\ }

\newcommand{\expec}[1]{\langle #1 \rangle}

\newcommand{\figref}[1]{Fig.~\ref{#1}}

\def\ie{i.e.\ }
\newcommand{\ig}[2]{\includegraphics[width=#1\tw,type=#2,ext=.#2,read=.#2]}

\def\lb{\label}

\def\mcal{\mathcal}

\def\non{\nonumber}

\def\ol{\overline}

\def\ra{\rightarrow}
\def\Ra{\Rightarrow}

\newcommand{\secref}[1]{Sec.~\ref{#1}}

\renewcommand{\th}[1]{^\tx{#1}}

\def\tit{\textit}

\newcommand{\tl}[1]{_\tx{#1}}

\def\tw{\textwidth}
\def\tx{\text}

\def\vt{\vert}

\def\wt{\widetilde}

\newcommand{\eq}[1]{\begin{align} #1 \end{align}}
\newcommand{\eqn}[1]{\begin{align*} #1 \end{align*}} 
\newcommand{\eqs}[1]{\begin{subequations}\begin{align} #1 \end{align}\end{subequations}}

\def\bar{\begin{array}}
\def\ear{\end{array}}

\def\bce{\begin{center}}
\def\ece{\end{center}}

\def\ben{\begin{enumerate}}
\def\een{\end{enumerate}}

\def\bfi{\begin{figure}[!tbh]\setcapindent{0em}\centering}
\newcommand{\bfiO}[1]{\begin{figure}[#1]\setcapindent{0em}\centering}
\def\efi{\end{figure}}

\def\bit{\begin{itemize}}
\def\eit{\end{itemize}}

\def\bqu{\begin{quote}}
\def\equ{\end{quote}}

\newcommand{\btblO}[2]{\begin{table}[#2]\setcapindent{0em}\centering\begin{minipage}{#1\tw}}
\newcommand{\btbl}[1]{\begin{table}[!tbh]\setcapindent{0em}\centering\begin{minipage}{#1\tw}}
\def\etbl{\end{minipage}\end{table}}

\def\btbr{\centering\vspace{.4em}\begin{tabular}}
\def\etbr{\end{tabular}}

\def\bseq{\begin{subequations}}
\def\eseq{\end{subequations}}

\def\bve{\begin{array}{l}}
\def\eve{\end{array}{l}}

\newcommand{\igS}[3]{%
\hspace{-.025\tw}\begin{minipage}[t]{#1\tw}\vspace{0em}\includegraphics[width=\linewidth,type=#2,ext=.#2,read=.#2]{#3}\end{minipage}%
\hspace{.03\tw}\begin{minipage}[t]{1\tw-#1\tw-.05\tw}}

\newcommand{\lbS}[1]{\lb{#1}\end{minipage}}

\makeatletter
  \newcommand{\myhref}[2]{\hyper@linkurl{#2}{#1}}
\makeatother

\makeatletter
\newcommand*\if@single[3]{%
  \setbox0\hbox{${\mathaccent"0362{#1}}^H$}%
  \setbox2\hbox{${\mathaccent"0362{\kern0pt#1}}^H$}%
  \ifdim\ht0=\ht2 #3\else #2\fi
  }
\newcommand*\rel@kern[1]{\kern#1\dimexpr\macc@kerna}
\newcommand*\widebar[1]{\@ifnextchar^{{\wide@bar{#1}{0}}}{\wide@bar{#1}{1}}}
\newcommand*\wide@bar[2]{\if@single{#1}{\wide@bar@{#1}{#2}{1}}{\wide@bar@{#1}{#2}{2}}}
\newcommand*\wide@bar@[3]{%
  \begingroup
  \def\mathaccent##1##2{%
    \if#32 \let\macc@nucleus\first@char \fi
    \setbox\z@\hbox{$\macc@style{\macc@nucleus}_{}$}%
    \setbox\tw@\hbox{$\macc@style{\macc@nucleus}{}_{}$}%
    \dimen@\wd\tw@
    \advance\dimen@-\wd\z@
    \divide\dimen@ 3
    \@tempdima\wd\tw@
    \advance\@tempdima-\scriptspace
    \divide\@tempdima 10
    \advance\dimen@-\@tempdima
    \ifdim\dimen@>\z@ \dimen@0pt\fi
    \rel@kern{0.6}\kern-\dimen@
    \if#31
      \overline{\rel@kern{-0.6}\kern\dimen@\macc@nucleus\rel@kern{0.4}\kern\dimen@}%
      \advance\dimen@0.4\dimexpr\macc@kerna
      \let\final@kern#2%
      \ifdim\dimen@<\z@ \let\final@kern1\fi
      \if\final@kern1 \kern-\dimen@\fi
    \else
      \overline{\rel@kern{-0.6}\kern\dimen@#1}%
    \fi
  }%
  \macc@depth\@ne
  \let\math@bgroup\@empty \let\math@egroup\macc@set@skewchar
  \mathsurround\z@ \frozen@everymath{\mathgroup\macc@group\relax}%
  \macc@set@skewchar\relax
  \let\mathaccentV\macc@nested@a
  \if#31
    \macc@nested@a\relax111{#1}%
  \else
    \def\gobble@till@marker##1\endmarker{}%
    \futurelet\first@char\gobble@till@marker#1\endmarker
    \ifcat\noexpand\first@char A\else
      \def\first@char{}%
    \fi
    \macc@nested@a\relax111{\first@char}%
  \fi
  \endgroup
}
\makeatother

\begin{document}

\title{Chebyshev Matrix Product State Impurity Solver\\
 for the Dynamical Mean-Field Theory}

\author{F. Alexander Wolf}
\affiliation{
Theoretical Nanophysics,
Arnold Sommerfeld Center for Theoretical Physics,
LMU M\"unchen,
Theresienstrasse 37,
D-80333 M\"unchen, Germany}
\author{Ian P. McCulloch}
\affiliation{
Centre for Engineered Quantum Systems,
School of Physical Sciences,
The University of Queensland,
Brisbane, Queensland 4072, Australia}
\author{Olivier Parcollet}
\affiliation{Institut de Physique Th{\'e}orique, CEA, IPhT, CNRS, URA 2306, F-91191 Gif-sur-Yvette, France}
\author{Ulrich Schollw\"ock}
\affiliation{
Theoretical Nanophysics,
Arnold Sommerfeld Center for Theoretical Physics,
LMU M\"unchen,
Theresienstrasse 37,
D-80333 M\"unchen, Germany}

\date{\today}

\begin{abstract}
We compute the spectral functions for the 
two-site dynamical cluster theory and for the 
two-orbital dynamical mean-field theory 
in the density-matrix renormalization group (DMRG) framework 
using Chebyshev expansions represented 
with matrix product states (MPS). We obtain 
quantitatively precise results at modest computational effort 
through technical improvements 
regarding the truncation scheme 
and the Chebyshev rescaling procedure. 
We furthermore establish the relation of the Chebyshev
iteration to real-time evolution, and discuss technical
aspects as computation time and
implementation in detail. 
\end{abstract}

\maketitle

\section{Introduction}

The dynamical mean-field theory (DMFT)\cp{metzner89,georges92,georges96,kotliar06} 
and its cluster extensions\cp{maier05} are among the most successful methods
to study strongly correlated electron systems in dimensions higher than one. The 
impurity problem within DMFT is usually solved with continuous-time 
quantum Monte Carlo (CTQMC) algorithms, \cp{gull11,rubtsov05,gull08i,werner06} the numerical
renormalization group (NRG) \cp{bulla08} or 
exact diagonalization (ED).\cp{caffarel94,granath12,lu14}
While CTQMC is computationally
feasible even for problems with many bands or a high number of 
cluster sites, it provides numerically exact results only on the 
imaginary frequency axis. 
Many experimentally relevant frequency-dependent quantities
like \eg the conductivity therefore can only  
be obtained via the numerically ill-conditioned 
analytical continuation. 
NRG, by contrast, solves the problem on the real frequency axis.  
But it badly resolves spectral functions at high energies 
and cannot treat DMFT calculations with more than e.g.\ two bands.  
The limiting factor for this is the exponential growth 
of the \tit{local} Hilbert space with the number of bands. 
Only recently, a reformulation of the mapping problem could 
avoid this exponential growth,\cp{mitchell14} but
it is still unclear whether this can be efficiently exploited 
in the context of DMFT.
ED faces the problem of a limited spectral resolution
due to the limited number of bath sites it can treat,
although recent publications could substantially improve that. \cp{granath12,lu14}


As the impurity problem of DMFT is one-dimensional, 
there has been a long-time interest to 
solve it using density matrix renormalization group (DMRG),\cp{white92,schollwock05,schollwock11} 
which operates on the class of matrix product states (MPS). DMRG 
features an unbiased energy resolution and shows no exponential growth
of the local Hilbert space with respect to the number of baths. It also works directly on the real-frequency axis, avoiding analytic continuation. 
The earliest DMRG approach to spectral functions, 
the Lanczos algorithm approach,\cp{hallberg95}
is computationally cheap, but does not yield 
high-quality DMFT results due to its intrinsic numerical instability.\cp{garcia04} 
Recent improvements using a fully MPS-based representation 
of this algorithm\cp{dargel12} are not sufficient to resolve this issue. \cp{wolf13iii}
The dynamical DMRG (DDMRG) approach\cp{kuhner99,jeckelmann02} 
yields very precise results for single-site DMFT 
on the real frequency axis,\cp{karski08,karski05,nishimoto04} 
but is computationally extremely costly and therefore not competitive 
with other impurity solvers for DMFT.

Recently, a new approach to spectral functions based on 
expansions in Chebyshev polynomials\cp{weisse06} represented 
with matrix product states (CheMPS)\cp{holzner11,braun13,tiegel14,ganahl14} was introduced by two  
of us in Ref.\ \onlinecite{holzner11}, 
which gave essentially the accuracy as the DDMRG approach 
at a fraction of the computational cost. At the same time, 
the availability of real-time evolution\cp{daley04,vidal04,white04} 
within time-dependent DMRG (tDMRG) and 
closely related methods generally also permits access 
to spectral functions by a Fourier transformation.\cp{white04} 
Both Chebyshev expansions (CheMPS)\cp{ganahl14} and 
tDMRG\cp{ganahl14i} were recently seen to be applicable to the solution of 
the DMFT. Both approaches are computationally cheaper 
than DDMRG and numerically stable. For the single-impurity 
single-band case, results on the real-frequency axis are excellent, but for more typical 
present-day DMFT setups involving clusters or multiple bands,  
results are not available in the case of Chebyshev expansions 
or do so far not reach the quality of the competing QMC and NRG methods 
in the case of real-time evolution.

In this paper, we push the application of CheMPS to DMFT further: 
(i) We solve the dynamical cluster approximation (DCA)\cp{maier05} 
for a two-site cluster and the DMFT for a two-band Hubbard model.  
The accuracy of the results for the latter case is better 
than those shown in Ref.\ \onlinecite{ganahl14i}, 
where the problem has been solved using tDMRG. 
(ii) We consider the experimentally relevant 
case of finite doping, which is significantly 
more complicated than the half-filled cases treated so far. 
(iii) We suggest a new truncation scheme for CheMPS, which allows 
to maintain the same error level at strongly reduced
computational cost.
(iv) We establish that the Chebyshev recurrence iteration can be interpreted
as a discrete real-time evolution.
(v) By comparing different methods to set up CheMPS, we obtain another
substantial increase in computation speed.
(vi) We discuss limitations of post-processing methods, 
which have been crucial to the success of DMRG as an DMFT impurity solver.

With these improvements, CheMPS immediately provides an efficient, 
precise and controlled way to solve DMFT problems with two baths (two-site clusters) 
on the real-frequency axis with feasible extensions to problems with more bands. 
The presentation proceeds as follows. After a general introduction 
to Chebyshev expansions of spectral functions in \secref{secCheb}, 
we move on to discuss its implementation in the approximate framework of MPS: 
in \secref{secCheMPS}, we present a new truncation scheme,
and in \secref{secOptCheb}, we discuss the mapping of the Hamiltonian 
to the $[-1,+1]$ convergence interval of Chebyshev polynomials, 
because this interacts non-trivially with efficient MPS calculations. 
\secref{secChebPost} treats the post-processing of Chebyshev 
moments obtained in the expansion. These improvements are then 
applied to various DMFT problems. As the case of the single-impurity 
single-band DMFT has been treated extensively in the literature and just serves 
as an initial benchmark, we move those results to the Appendix.
In the main text, we give examples for the relevance of our 
improvements to CheMPS by solving a two-site DCA in \secref{secVBDMFT} 
and a single-site two-orbital DMFT in \secref{secTIAM}.
Technical details of these calculations are again found in 
the Appendix. \secref{secConc} concludes the paper.

\section{Chebyshev expansion of spectral functions}
\label{secCheb}
In this Section, we establish notation and explain the general ideas 
behind Chebyshev expansions of spectral functions.
The zero-temperature single-particle Green's function associated with a
many-body hamiltonian $H$ is
\eq{
  \quad G(\w) = \bra{E_0} c \frac{1}{\w+i0^+ - (H - E_0) } c\dag \ket{E_0},
}
where $c\dag$ creates a particle in a particular quantum state and 
$\ket{E_0}$ is the ground state with energy $E_0$.
The spectral function $A(\w) = -\frac{1}{\pi} \tx{Im}\,G(\w)$ reads
\eq{ \label{eqRhoH}
  A(\w) & = \, \bra{E_0} c\, \delta(\w - (H - E_0)) c\dag \ket{E_0} \non \\
	    &=  \sum_n  W_n \delta(\w - (E_n-E_0)),
}
with weights $W_n=\vt\bra{E_n} c\dag \ket{E_0}\vt^2$.
If evaluated exactly in a finite system, $A(\w)$ is a comb of delta peaks,
which only in the thermodynamic limit becomes a smooth function 
$A\tl{lim}(\w)$. If evaluated in an approximate way that
averages over the finite-size structure of $A(\w)$, it is possible to
extract $A\tl{lim}(\w)$ also from a sufficiently big finite-size system. Among various 
techniques that provide such an approximation,\cp{lin13} the
most popular one is the definition of a \tit{broadened}
representation of $A(\w)$
\eq{ \label{eqRhoEta}
  A_\eta(\w) & = \sum_{n} W_n h_\eta(\w-E_n) 
} 
where the broadening function $ h_\eta(\w-E_n)$ is given by the Gaussian kernel 
\eq{ \label{eqRhoEtaGauss}
h_\eta(x) = \frac{1}{\sqrt{2\pi}\eta} e^{-\frac{x^2}{2\eta^2}}.
} 
Besides the Gaussian kernel, a Lorentzian kernel  
\eq{ \label{eqRhoEtaLorentz}
h_\eta(x) = \frac{\eta}{\pi} \frac{1}{x^2+\eta^2}
} 
is often implicitly used 
as it emerges automatically when computing the spectral function 
$A_\eta = -\frac{1}{\pi} \tx{Im}\,G(\w+i\eta)$
from the shifted Green's function $G(\w+i\eta)$. 
In general, $A_\eta(\w)$ is indistinguishable from
$A\tl{lim}(\w)$ if the latter has no structure on a scale smaller than $\eta$.

An efficient way to generate the 
broadened version $A_\eta(\w)$ of $A(\w)$ 
is via iterative expansions in orthogonal polynomials. 
Historically most frequently used in this context is the Lanczos algorithm, which
is intrinsically numerically unstable, though. By contrast,  
expansions in Chebyshev polynomials 
can be generated in a numerically stable way. 
As they haven't been used much in either the DMRG or DMFT community so far, 
we briefly introduce them based on Ref.~\onlinecite{weisse06}.

\subsection{General implementation}

The Chebyshev polynomials of the first
kind $T_n(x)$ can be represented explicitly by
\eq{
  T_{n}(x) = \cos\left(n \arccos(x)\right)   \label{eqTn}
}
or generated with the recursion
\eq{ \label{eqRec}
  T_{n}(x) = 2xT_{n-1}(x) - T_{n-2}(x), \quad
  T_{0} = 1, \quad T_{1} = x,
}
which is numerically stable if $\abs{x} \leq 1$. Chebyshev 
polynomials are orthonormal with respect to the weighted 
scalar product
\eqs{
 \int_{-1}^{1} dx\, w_n(x) T_m(x) T_n(x) & = \d_{nm}, \\
 w_n(x) & = \frac{2-\d_{n0}}{\pi \sqrt{1-x^2}}.     \label{eqMeasure}
}
Any sufficiently well-behaved function $f(x)\vert_{x\in[-1,1]}$ can be expanded in Chebyshev polynomials
\eqs{ \label{eqChebExp}
  f(x) &=
  \sum_{n=0}^{\infty} w_n(x) \mu_{n} T_{n}(x), \\
  \mu_n &= \int_{-1}^{1} dx f(x) T_n(x),   \label{eqScalarProd}
}
where the definition of the so-called \tit{Chebyshev moments} $\mu_n$ via
the \tit{non-weighted} scalar product follows when applying $\int_{-1}^{1}dx\, T_m(x) \dots$
to both sides of  \eqref{eqChebExp}.

If $f(n)$ is smooth, the envelope of $\mu_n$ decreases at least exponentially to zero 
with respect to $n$; if $f(n)$ is the step function, the envelope decreases algebraically; and 
if $f(n)$ is the delta function, the envelope remains constant. \cp{boyd01} 
For a smooth function, the
truncated expansion $f_N(x)=\sum_{n=0}^{N} w_n(x)\mu_n T_n(x)$ 
therefore 
approximates $f(x)$ very well if $N$ is chosen high enough.
But for the delta function,
any truncated expansion yields an approximation 
with spurious (Gibbs) oscillations.
A controlled damping scheme for the oscillations, the so-called
\tit{kernel polynomial approximation} (KPM), can be obtained 
with a simple modification of the Chebyshev expansion,
\eqs{ 
f_{N}\th{kernel} &=  \sum_{n=0}^{N} w_n(x) g_n \mu_{n} T_{n}(x),\\
  g_{n} &= \frac{(N-n+1)\cos\frac{\pi n}{N+1} +
  \sin\frac{\pi n}{N+1}\cot\frac{\pi}{N+1}}{N+1} ,   \label{eqJack}
}
where $g_n$ is the so-called Jackson kernel that leads to a very good Gaussian 
approximation $h_{\eta(x)}(x)$ with $x$-dependent
width $\eta(x)=\sqrt{1-x^2}\,\pi/N$ of the delta function,
and hence directly leads to \eqref{eqRhoEtaGauss}. 

In the case of the spectral function \eqref{eqRhoH}, 
one aims at an expansion of 
a superposition of delta functions. This 
can in practice often be done without damping: 
When expanding \eqref{eqRhoH}
in Chebyshev polynomials, the integration in \eqref{eqScalarProd}
averages over the delta-peak 
as well as over the finite-size peak structure of $A(\w)$. If the 
weights $W_n$ vary slowly on the scale of the spacing of
finite-size peaks, the sequence $\mu_n$ 
approaches zero as soon as the characteristic 
form of this slow variation is resolved. The value of $n$
at which this \tit{pseudo}-convergence occurs is the one that
resolves the spectral function in the thermodynamic limit $A\tl{lim}(\w)$,
provided that  $A\tl{lim}(\w)$ has no structure on a smaller scale than the 
spacing of finite-size peaks.
Only for much higher values of $n$, 
the Chebyshev moments start deviating from zero again to
then oscillate forever, resolving first the finite-size structure of $A(\w)$
and finally the delta-peak structure. 
Therefore, if one can generate the sequence up to \tit{pseudo}-convergence, then there is no need for 
Jackson damping.

\subsection{Operator valued Chebyshev expansion}

In order to expand the spectral
function \eqref{eqRhoH}, one usually introduces 
a rescaled and shifted version of $H$ in order to map its spectrum 
into the interval $[-1,1]$, where Chebyshev polynomials are bounded 
and have a stable recursion relationship,
\eq{
  H' = \frac{H-E_0+b}{a}, \quad \w' = \frac{\w+b}{a}.       \label{eqShiftScale}
}
Obviously, there is a lot of leeway in the choice of $a$ and $b$, 
which will be found to have large implications for CheMPS (\secref{secOptCheb}). Generally, 
\eq{
  A(\w) & =  \frac{1}{a} A'\big(\tfrac{\w+b}{a}\big), \tx{ where } \non  \\
  A'(\w') & = \bra{t_0} \delta(\w' - H') \ket{t_0}, \quad \ket{t_0} = c\dag\ket{E_0}.  \label{eqAscaled} 
}
Expanding $A'(\w')$ in Chebyshev polynomials yields the moments 
\eq{ 
  \mu_n 
  & = \int_{-1}^{1} d\w' \bra{t_0} \d(\w' - H') \ket{t_0} T_n(\w') \non\\
  & = \sum_i\int_{-1}^1 d\w' \bra{t_0} \d(\w'-E_i') T_n(\w') \ket{E_i}\bra{E_i}t_0\rangle  \non\\  
  & = \ip{t_0}{t_n},  \quad \ket{t_n} = T_n(H') \ket{t_0}, \label{eqMuH}
}
Inserting the recursive definition \eqref{eqRec} of $T_n(H')$ in the definition of $\ket{t_n}$
one obtains a practical calculation scheme for the power series expansion of $T_n(H')$
\eqs{ 
  \ket{t_{n}}  & = 2 H' \ket{t_{n-1}} - \ket{t_{n-2}}  ,     \label{eqtRec}  \\
  \ket{t_{0}} & = c\dag\ket{E_0},\qquad \ket{t_{1}} = H' \ket{t_{0}}.
}
One can double the expansion order with the following relation\cp{weisse06}
\eqs{
\ol{\mu}_{2n-1} & = 2 \ip{t_n}{t_{n-1}} - \mu_1,\\
\ol{\mu}_{2n} & = 2 \ip{t_n}{t_{n}} - \mu_0,
}
but has to be aware of the fact that moments computed
this way are more prone to numerical errors.\cp{holzner11}

\subsection{Retarded fermionic Green's function}
\label{secAnalytic}
In the case of fermionic problems, as encountered in DMFT, 
an additional technical complication comes up.
The spectral representation of the fermionic 
retarded Green's function 
is the sum of its particle and hole parts
\eq{
A(\w) & = A^>(\w) + A^<(-\w) , \non\\
A^>(\w) & = \bra{E_0} c_0 \, \d(\w - (H - E_0)) c_0\dag \ket{E_0} , \non \\
A^<(\w) & = \bra{E_0} c_0\dag\, \d(\w - (H - E_0)) c_0 \ket{E_0}.
}
As $A^{\lessgtr}(\w)$ have steps at $\w=0$, 
their representation in terms of smooth polynomials
is notoriously ill-con\-di\-tio\-ned. One should therefore try to represent the smooth function
$A(\w)$ by a single Chebyshev expansion: Allowing for two different
rescaling prescriptions, one has
\eqs{
  A^>(\w) & = \frac{1}{a_1} \sum_n w_n(\w_1'(\w)) \mu_n^> T_n(\w_1'(\w)) \label{eqMuGtr}   \\
  A^<(-\w) & = \frac{1}{a_2} \sum_n w_n(\w_2'(-\w)) \mu_n^< T_n(\w_2'(-\w))
}
In order to write $A(\w)$ in terms of 
a single Chebyshev expansion, one can use the symmetries 
$T_n(x) = (-1)^n T_n(-x)$ and
$w_n(x)=w_n(-x)$. These restrict 
the rescaling parameters via $\w_1'(\w)=-\w_2'(-\w)$ 
to  $a_1=a_2=a$ and $b_1=-b_2$.
Making the particular choice $b_1=b_2=b=0$
hence defines a common expansion via\cp{ganahl14}
\eq{
A(\w) & = \frac{1}{a} \sum_n w_n(\tfrac{\w}{a}) (\mu_n^> + (-1)^n \mu_n^<) T_n(\tfrac{\w}{a}).
}

Although $b=0$ provides one with 
a controlled treatment of the step function, it comes at
the price of a loss in computational speed.
We will compare advantages and disadvantages of two practical shifting possibilities
($b=0$ and $b=-a$) in detail in \secref{secOptCheb}.

\section{Matrix Product implementation}
\label{secCheMPS}

So far everything has been general, or it was somehow assumed 
that all calculations can be carried out exactly, 
which meets severe limitations in computational practice. 
Representing Chebyshev states $\ket{t_n}$ with matrix 
product states (MPS)\cp{holzner11} 
enables more efficient computations than in an exact representation, 
as the size of the effective Hilbert space can be tremendously reduced. 
As an MPS is usually only an approximate representation of a 
strongly correlated quantum state, the issue of optimal compression, 
i.e.\ the representation of a quantum state as an MPS using 
finite-dimensional matrices with a minimal loss of accuracy (information), 
is crucial. Here, we argue in the following that instead of controlling the maximal matrix
dimension,\cp{holzner11,ganahl14,tiegel14} one should rather control 
the cumulated truncated weight (a proxy measure of the loss of accuracy), 
allowing for more efficient and more controlled calculations of Chebyshev moments.

\subsection{Adaptive matrix dimension}
\label{secAdaptm}

\begin{figure}
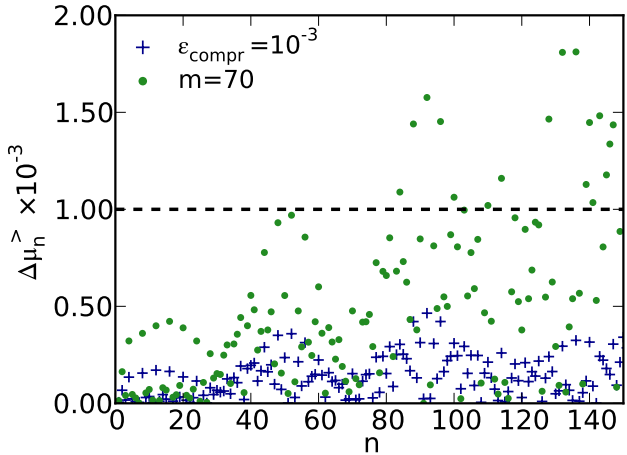

\ig{0.48}{pdf}{muerr}
\caption{(Color online) Error of Chebyshev moments $\mu_n^>$ (as they appear in \eqref{eqMuGtr}), 
computed as $\Delta\mu_n^> = \abs{\mu_n^> - \tilde\mu_n^>}$, where
$\tilde\mu_n^>$ is obtained with a quasi-exact calculation 
with high matrix dimension $m=200$.
If one fixes the matrix dimension $m$, the error steadily increases. If, instead, one fixes the 
cumulated truncated
weight $\ve\tl{compr}$, the error remains approximately constant and does not accumulate. This is the procedure followed in this paper.
As here, $\abs{\ket{t_0}}^2=1$, $\varepsilon\tl{compr}$ equals the upper error bound
of a single compression step. Results shown are for the spectral function of 
the half-filled single-impurity Anderson model (SIAM) (\appref{secSIAM}) with 
semi-elliptic density of states of half-bandwidth $D$, interaction 
$U=2D$, represented on a chain 
with $L=40$ lattice sites. This is equivalent to considering  
the local density of states at the first site of a fermionic 
chain with constant hopping $t=D/2$ and an interaction of $U=4t$ 
that acts solely at the first site. 
}
\label{figMuerr}
\end{figure}

If one follows through the recursive scheme for Chebyshev vectors, 
one starts out from a ground state, which we may assume has 
been obtained by a standard DMRG (MPS) calculation to 
extremely high precision, this means that 
an optimally compressed starting MPS is 
available where matrices have some computationally feasible 
dimension at very small loss of accuracy compared to the 
exact starting state. This, in turn, yields an extremely 
precise starting Chebyshev state $|t_0\rangle$. Now, 
in each step of the recursion \eqref{eqtRec}, 
one applies $H'$ and subtracts a preceding Chebyshev state. 
As is well-known for MPS, the application of $H'$ 
(and to a lesser extent the subtraction) lead to a drastic 
increase in matrix dimension, which necessitates a state compression (Sec.\ 4.5 of Ref.\ \onlinecite{schollwock11})
of the new Chebyshev state $\ket{\wt t_n}$ 
to a computationally manageable state $\ket{t_n}$ with smaller matrix dimension $m$,
which generates the error $\delta$
\eq{
\mu_n & = \bra{t_0} \wt t_n \rangle  = \bra{t_0} t_n \rangle \pm \delta, \\
\ket{\wt{t_n}} & =  2H' \ket{t_{n-1}} - \ket{t_{n-2}}, \non \\
\delta^2 & = \abs{\bra{t_0} (\ket{\wt{t_n}} - \ket{t_n})}^2 \non\\ 
		 & < \abs{\ket{t_0}}^2 \abs{\ket{\wt{t_n}} - \ket{t_n}}^2 < \abs{\ket{t_0}}^2  \varepsilon\tl{compr}(m).  \non
}
Here, we used the upper error bound\cp{verstraete06} provided by the 
cumulated truncated weight $\varepsilon\tl{compr}(m)$
\eq{
\abs{\ket{\wt{t_n}} - \ket{t_n}}^2 \leq \varepsilon\tl{compr}(m) = \sum_{i=1}^{L-1} \epsilon_i(m), \label{eqDiscardedweight}
}
where $\epsilon_i(m)$ is the sum over the discarded reduced density-matrix 
eigenvalues per bond and the sum over $i$ is over all bonds. 
This error bound for a single step of the recursion unfortunately
does \tit{not} provide a statement about 
the total error that accumulates over
all compression steps in preceding Chebyshev recursion steps. Still, we experienced 
that the numerical stability of the Chebyshev  
recursion rather leads to a helpful
compensation of errors of single recursion steps.
\figref{figMuerr} shows that the {\em total} error 
stays at the order of the error
of a {\em single} step $\abs{\ket{t_0}}^2 \varepsilon(m)$ 
also for high iteration numbers $n$. 
In the case in which one fixes
the matrix dimension $m$, \figref{figMuerr}
shows a steady, uncontrolled increase of the total error.
This is particularly undesirable in view of the 
desired post-processing of Chebyshev moments (\secref{secChebPost}). 

Another possibility would be to fix the local 
discarded weight $\e_i(m)$ as defined
in \eqref{eqDiscardedweight}. But this  
does in general \tit{not} lead to a viable computation scheme for 
impurity models: In the simplest and most-employed chain representation of impurity models,
the impurity site is located at an edge of the chain. 
Fixing the same value for $\e_i(m)$ for all bonds 
then leads to extremely high matrix dimensions
in the center of the chain, i.e.\ in the center of the bath,
where entanglement for systems with open boundary conditions is maximal.
The relevant entanglement, by contrast, is the one 
between the impurity site and the bath. This becomes clear
when noticing that upon projecting the Chebyshev state $\ket{t_n}$
on $\ket{t_0}$ to compute $\mu_n$, only correlations
with respect to the \tit{local} excitation $c\dag\ket{E_0}$
are measured. The high computational effort of high matrix dimensions
that follows when faithfully representing entanglement \tit{within} the bath,
is therefore in vain. For geometries with the impurity
at the center, like the two-chain
geometry used for the two-bath problems in this paper, 
the preceding argument is not valid. An inhomogeneous 
distribution of matrix dimensions with high values 
at the center and low values at the boundaries 
is \tit{a priori} consistent with open boundary conditions. 
This distribution can therefore be achieved by fixing a constant value
for $\e_i(m)$ for each bond. Another possible truncation scheme could 
be obtained by using an estimator for the correlations of the impurity with the bath,
which then fixes the matrix dimensions as a function of bonds $m(i)$ (distance to the impurity). 
Both approaches constitute possible future refinements. For simplicity, in this paper,
we consider the truncation scheme that fixes a constant value of $m$ based on the
cumulative truncated weight.

\subsection{State compression}
\label{secCompress}

During the repeated solution of \eqref{eqtRec} we monitor the truncated weight $\varepsilon\tl{compr}$. 
If  $\varepsilon\tl{compr}$ exceeds a certain threshold of the order of $10^{-4}$ to $10^{-3}$, 
we slightly increase the matrix dimension $m$, and repeat the compression. 
For the first compression step we take as an initial
guess the previous Chebyshev state $\ket{t_{n-1}}$. For 
repeated compression steps we take as an initial guess the state
of the previous compression step. 
It turns out that in practice one almost never faces repeated compressions,
which gains one approximately a factor 2 in computation speed
compared to the error monitoring of Ref.~\onlinecite{holzner11}: in Ref.~\onlinecite{holzner11}, the authors keep the matrix dimension fixed and 
variationally\cp{schollwock11} compress an exact representation of
the right hand side of \eqref{eqtRec} for fixed $m$ 
by repeated iterations (``sweeps'') until the error
\eq{
\left\vt 1 - \frac{\ip{t_n'}{t_n}}{ \vt\vt\,\ket{t_n'}\,\vt\vt\; \vt\vt\,\ket{t_n}\,\vt\vt }  \right\vt,
}
drops below a certain threshold. Here, $\ket{t_n'}$ denotes the state before
a sweep, and $\ket{t_n}$ the state after a sweep. 
This error measure is not related to the factual error of Chebyshev moments,
for any but the first sweep. Its monitoring is costly to compute 
and leads to at least two compression sweeps.

\section{Optimal Chebyshev setup}
\label{secOptCheb}

One can generally state that 
the effectiveness of the MPS evaluation of the Chebyshev recursion \eqref{eqtRec} 
for a certain system is unknown \tit{a priori} but must be experienced by
observing how strong entanglement in the Chebyshev vectors, 
and therefore matrix dimension $m$ 
needed for a faithful representation grows 
as compared to the speed of convergence of $\mu_n$.  
For very high iteration numbers one will always
reach a regime in which matrix dimensions have 
grown so much that further calculations become too expensive computationally.
This is known from tDMRG as \tit{hitting an exponential wall}
and defines an \tit{accessible time scale}, or in our case,
an accessible expansion order.  
In the case of the computation of Chebyshev moments, the 
accessible time scale strongly depends on the choice of 
the shifting parameter $b$, which leads us to consider the 
two cases $b=0$ and $b=-a$.

Comparing these cases, one finds a much slower speed of convergence
of the Chebyshev moments in the case $b=0$ than in the case $b=-a$. 
Putting that differently: per fixed amount of entanglement 
growth (application of $H$ in one step of \eqref{eqtRec}),
much less information about the spectral function is extracted
in case $b=0$ than in case $b=-a$.  
Independent of that, one finds that the advantage of the choice $b=0$ 
to provide one with an analytic expression for $A(\w)$ in terms of a 
single Chebyshev expansion (\secref{secAnalytic}) can be detrimental.
We therefore need to study both cases in more detail. 

\subsection{No shift: $b=0$}
\label{secb0}

If choosing $b=0$, one can derive a scaling property of Chebyshev moments
that simplifies extracting the thermodynamic limit as well as 
the examination of computational performance.

The spectral function of a one-particle operator $A(\w)$
is non-zero only in the vicinity of the groundstate energy $\w=0$,
up to a distance of the order of the single-particle bandwidth $W\tl{single}$.
The rescaled spectral function $A'(\w')$ is non-zero 
up to a distance of $W\tl{single}/a$ 
from $\w'=0$. For all rescaling parameters $a$ 
that have been proposed up to now,\cp{weisse06,holzner11,ganahl14}
one has $W\tl{single}/a < \frac{1}{2}$.
Usually $W\tl{single}/a$ is much smaller than the upper bound $\frac{1}{2}$. 
As $\arccos(x) = \pi/2 - x - x^3/6 + \dots$ is well
approximated by its linear term already for $\abs{x} < 0.5$, 
Chebyshev polynomials \eqref{eqTn} 
behave like a shifted cosine function
in the region where  $A'(\w')$ is non-zero. 
The expansion of $A'(\w')$ in Chebyshev polynomials 
is therefore essentially equivalent to a Fourier expansion. This means
that the iteration number $n$ of the Chebyshev expansion 
has the same meaning as a discrete propagation time,
the evolution of which is mediated by simple applications of $H$ instead
of the ordinary continuous time propagation $e^{-iHt}$. To answer the question of whether
an ordinary time evolution\cp{ganahl14i} 
is more effective in generating information
about the spectral function, one has to study the entanglement
entropy production of repeated applications of $H$ compared 
to the one of $e^{-iHt}$. The following results are first steps in this direction.

\begin{figure}
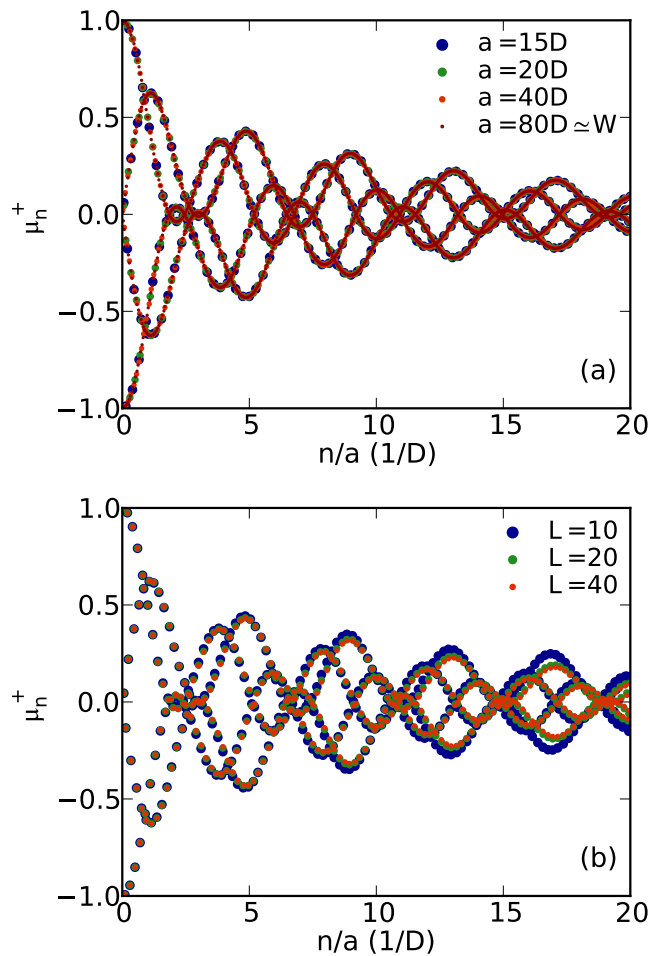

\ig{0.48}{pdf}{scalingmu}
\ig{0.48}{pdf}{scalingfinitesize}
\caption{(Color online) Panel (a): Chebyshev moments $\mu_n^>$ vs $n/a$ for fixed system size
and different values of $a$ and $b=0$. Except for a different total number of points, the rescaled moments 
all lie on the line obtained when $a\ra\infty$ and $n/a$ becomes continuous.
Here, we study the half-filled SIAM (\appref{secSIAM}) with semi-elliptic density 
of states of half-bandwidth $D$ and $U=2D$, 
represented on a chain with length $L=80$. The full many-body bandwidth is $W\simeq80D$.
Panel (b): Chebyshev moments for different system sizes $L$. Except for 
the system size and the scaling parameter, parameters are the same as in panel (a). 
Here all calculations were done with a rescaling constant of $a=20D$. 
For low values of $n$, the results for different system sizes are virtually indistinguishable.
For higher values of $n$, moments start to disagree 
as finite-size features start to be resolved. The $L=80$ and the 
$L=40$ results would be indistinguishable in this plot. 
}
\label{figScaling}
\end{figure}

In discrete time evolution, the rescaling of the frequency
directly translates to an inverse scaling of time. Considering two calculations
of Chebyshev moments, one for $\mu_n^{(1)}$ performed with $H'$ 
and another for $\mu_n^{(a)}$ 
performed with $H'/a$,
one therefore has the simple approximate relation
\eq{ \label{eqScalingApprox}
\mu_n^{(1)} & \sim \bra{t_0}\cos(n H')\ket{t_0}  \non\\
   &  = \bra{t_0}\cos(a n H'/a)\ket{t_0}  \sim \mu_{na}^{(a)} .
}
This means that if rescaling with $a$, one has to compute $a$ times more
Chebyshev moments than in the case without rescaling. 
An exact version of statement \eqref{eqScalingApprox}
is given in \eqref{eqScaling} in \appref{secScaling}. \figref{figScaling}(a) illustrates 
the scaling property \eqref{eqScalingApprox} for a 
system of fixed size. 
\paragraph{Extracting the thermodynamic limit.}

One direct application of the scaling property \eqref{eqScalingApprox}, 
lies in the study of the thermodynamic limit by comparing systems
of increasing size $L$. For low values of $n$, even small
systems have the same Chebyshev moments as
in the thermodynamic limit.
Finite-size features are averaged out in 
the integral \eqref{eqScalarProd} as long as 
$T_n(x)$ oscillates slowly enough. $T_n(x)$ 
oscillates $n$ times on $[-1,1]$. 
An $N$th order Chebyshev expansion therefore resolves
features on the scale $2/N$, 
which on the original energy scale is $2a/N$.
Finite-size oscillations appear at a spacing of $W\tl{single}/L$, 
where $W\tl{single}$ is the single-particle bandwidth. 
Equating resolution  with the spacing of finite-size oscillations
\eq{
2a/N\tl{finsize}  = W\tl{single}/L,
}
gives the expansion order  $N\tl{finsize}$ at which 
finite-size features are first resolved. 
\figref{figScaling}(b) illustrates these statements 
by comparing Chebyshev moments
computed for different system sizes.

\paragraph{Optimizing computation time.} 

\begin{figure}
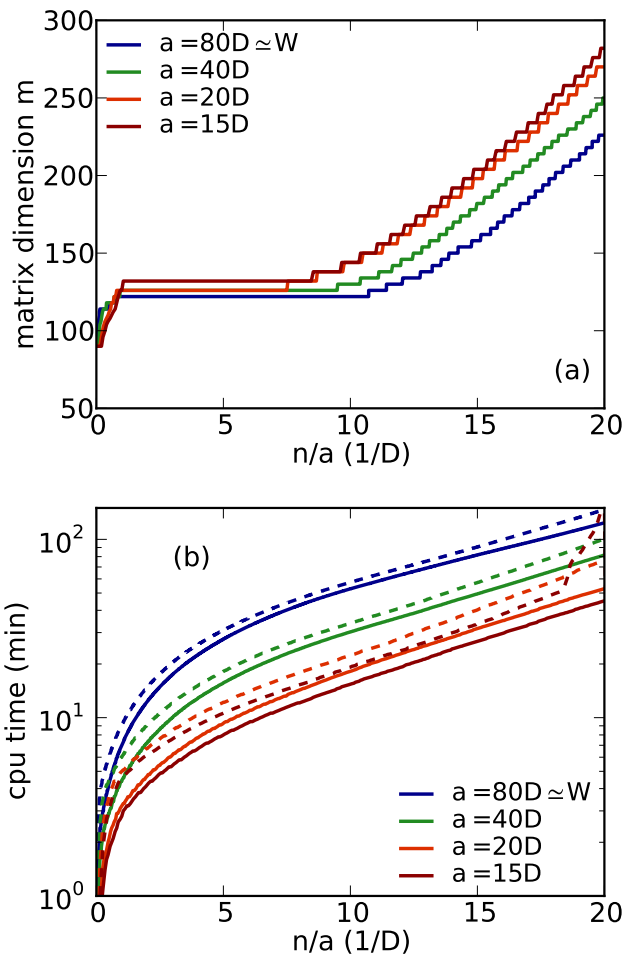

\ig{0.48}{pdf}{scalingmatrixdim}
\ig{0.48}{pdf}{scalingcput}
\caption{(Color online) Performance of the adaptive matrix dimension algorithm 
(\secref{secAdaptm})
for the example described in the caption of \figref{figScaling}.
Panel (a): Adaption of matrix dimensions for different rescaling 
factors, fixing a truncation error of $\varepsilon\tl{compr}=10^{-3}$. 
Panel (b): Computer time needed to generate the same amount
of information for different scalings running on a single-core
2.0$\,$GHz workstation. Solid lines: fixing 
a truncation error of $\varepsilon\tl{compr}=10^{-3}$. 
Dashed lines: $\varepsilon\tl{compr}=5\times10^{-4}$. 
The iteration number where the 
irregular behavior of the dashed line for $a=15 D$ 
starts corresponds to the point where numerical errors 
render the Chebyshev recursion unstable. Note that while small $a$ leads to the 
largest matrix sizes, which is costly in MPS, the overall cost of CPU 
time nevertheless is lowest, as a smaller expansion order is needed.}
\label{figcput}
\end{figure}

\figref{figcput} shows how 
computation time depends on the rescaling constant 
$a$ for the example 
of the moments shown in \figref{figScaling}(a). As already qualitatively 
stated previously \cp{holzner11,ganahl14}, one observes that upon
using a lower value of $a$ computation time is reduced. 
In all cases, computation time diverges exponentially (\figref{figScaling}(b)).
Note that rescaling with a higher value of $a$ allows to 
compute at smaller matrix dimensions. Note further that if choosing
$a$ too small, numerical errors can render the recursion \eqref{eqtRec}
unstable. In contrast to common belief, it is possible to use much smaller
values of $a$ than the full many-body bandwidth. 
Achieving even smaller values of $a$ can be done with 
the so-called \tit{energy truncation}\cp{holzner11}, but 
after several tests, we did not find this to lead to an effective 
speed-up of calculations. We therefore discard it in our calculations 
as a source of additional tuning parameters. 
We have also tested the idea of \ct{ganahl14} to map the spectrum
of $H$ into $[-1,1]$ via $1-\exp(\beta H)$. 
The idea might be worth to study in more detail, 
but again, we could not gain any performance improvement
over a simple rescaling procedure.

\subsection{Shifting by $b=-a$.}
\label{secba}

The choice $b=-a$ in \eqref{eqShiftScale} makes
an analytic expression of the complete spectral function
$A(\w)=A^+(\w)+A^-(-\w)$ in terms of a single
Chebyshev expansion impossible, but has beneficial effects
on the computation time. This is to be 
understood in the following sense: Due to the 
increased oscillation frequency of $T_n(x)$ 
close to the interval boundaries of $[-1,1]$, the integral
\eqref{eqChebExp} extracts much more information about
the spectral function in the vicinity of these boundaries.  
This is reflected \eg in the fact that 
the width of the Gaussian obtained by the kernel 
polynomial expansion approaches zero close the 
interval boundaries of $[-1,1]$ (see the discussion below \eqref{eqJack}).
It is therefore desirable to shift the relevant
part of the spectral function, the part slightly above
the Fermi edge, to match the left boundary $-1$. This is achieved
by the choice $b=-a$. In practice, one adds a small
correction $a \e$, $\e\sim10^{-3}$, to avoid problems
with the diverging weight function $w_n(x)$ in \eqref{eqMeasure}.

Another advantage of the $b=-a$ setup is that one can use a smaller
scaling constant $a$ than in the $b=0$ setup. The Chebyshev
iteration becomes unstable when the iteration number $n$ 
becomes so high that $\ket{t_n}$ has accumulated erroneous 
contributions from eigen states with eigen energies $E_n'=(E_n-E_0+b)/a>1$. 
For fixed $a$, the additional subtraction in the $b=-a$ setup ensures that 
the instability appears for a higher iteration number than in the $b=0$ setup.
Therefore, the $b=-a$ setup allows smaller values of $a$. 
We finally note that the choice $b=-a$ is equivalent to the choice suggested by \ct{weisse06}, 
if one rescales with the full many-body bandwidth $a=W$. 
In this case, the computation can be carried out to arbitrarily high order
and will never become unstable. In the $b=0$ setup, one would have
to choose $a=2W$ to reach arbitrarily high expansion orders.
\begin{figure}
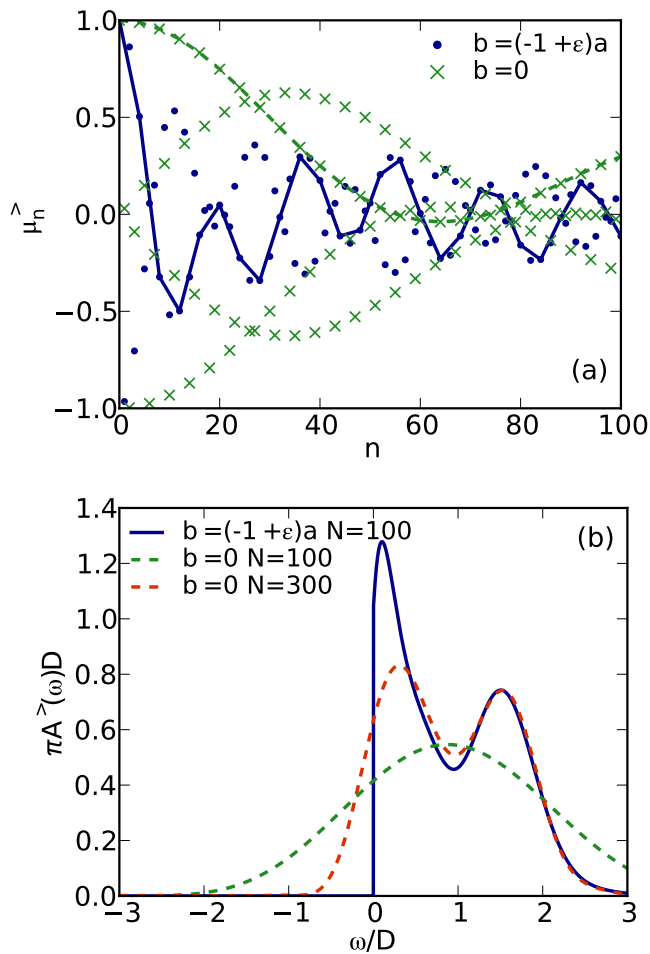

\ig{0.48}{pdf}{shift_mom}
\ig{0.48}{pdf}{shift_rho}
\caption{(Color online) Local particle density of states of the half-filled 
SIAM (\appref{secSIAM})
with semi-elliptic density of states of half-bandwidth $D$. 
$L=40$, $U=2D$ and $a=30 D$ in all cases.
Panel (a): Chebyshev moments. Lines connect every 4th moment
and by that reveal the relevant slow oscillation. They are a guide to the eye.
Panel (b): Corresponding spectral functions evaluated using Jackson damping
\eqref{eqJack}. The $b=0$ calculation
requires three times more iterations than the 
$b=-a$ calculation
to resolve the right Hubbard peak with the same  
resolution. In this case the central peak is still
much better resolved for $b=-a$.
}
\label{figShift}
\end{figure}

In \figref{figShift}(a), we plot Chebyshev moments 
for both types of shifts $b=0$ and $b=-a$.
The moments obtained for $b=0$ show a slow structureless oscillation
whereas the moments obtained for $b=-a$ show a much faster oscillation.
\figref{figShift}(b) shows that upon using the same rescaling
constant $a$ and the same expansion order $N=100$, which 
leads to very similar entanglement growth,  
both shift types differ strongly in the achieved resolution. 
To resolve at least the right Hubbard peak 
with a $b=0$ calculation at the resolution of $b=-a$ calculation,
one needs $N=300$ moments. As computation time
increases exponentially (\figref{figcput}(b)) with respect to expansion order $N$ in 
both cases, this difference is highly relevant. 

We apply both setups, $b=0$ and $b=-a$, 
to the benchmark test of the DCA in \secref{secVBDMFT}, 
and find a significant speed-up for $b=-a$ at a small
loss in accuracy. Previously,\cp{ganahl14} only $b=0$ has
been considered for the solution of the DMFT.

\section{Post-processing moments}
\label{secChebPost}

Whereas Jackson damping \eqref{eqJack} can be seen as one possibility
to post-process Chebyshev moments in order to achieve 
uniform convergence even for the truncated Chebyshev expansion
of a delta function, there is another, fundamentally different approach.

The computation of the Chebyshev
moments becomes very costly for high iteration numbers.
In the case in which Chebyshev moments start
to follow a regular pattern when $n$ exceeds a certain threshold,
it is possible to continue this pattern to infinity, and one
can avoid the costly computation of moments.
Consider the typical example in which the spectral function  
is a superposition of Lorentzians (quasiparticle peaks) 
and of a slowly varying background density.
As for low values of $n$, 
$T_n(x)$ extracts information via \eqref{eqScalarProd} only 
about the slowly varying background density,
while for high values of $n$,  $T_n(x)$ extracts information
only about the sharp and regular Lorentzian structures,  
$\mu_n$ starts to follow a regular pattern for high 
numbers of $n$. For a sum of Lorentzians, with weights $\alpha_i$, 
widths $\eta_i$, and positions $\w_i$, this pattern can be 
obtained analytically:
\eq{
  A\tl{Lor}(\w) & = \sum_i \alpha_i \frac{\eta_i}{\pi} \frac{1}{(\w-\w_i)^2+\eta_i^2}, \non\\
 \Ra \quad \mu_n   & \simeq \sum_i \alpha_i \cos(n(\w_i-\frac{\pi}{2})) e^{-n\eta_i}, \label{eqMuLo}
}
as shown in \appref{secMuLorentzGauss}. If one recalls (\secref{secOptCheb}) that the Chebyshev 
recursion corresponds to a discrete time evolution if choosing $b=0$, 
the result of \eqref{eqMuLo} could have been anticipated.

\begin{figure}
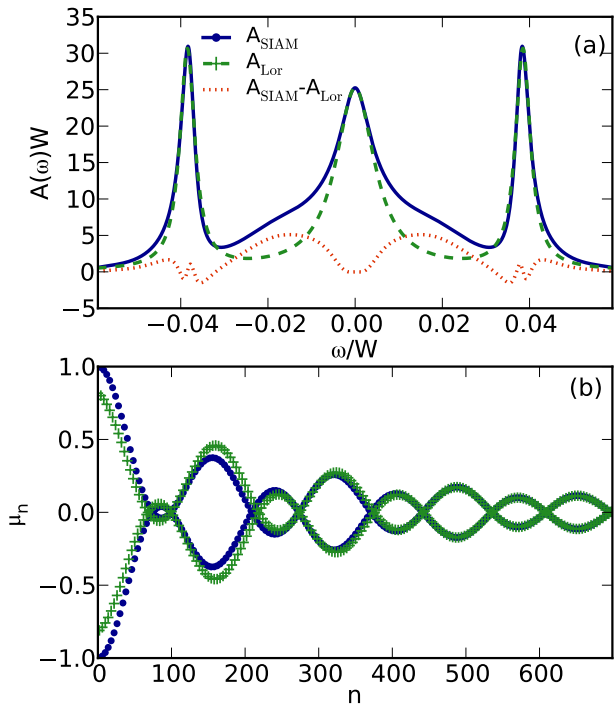

\ig{0.48}{pdf}{chebmomLo}
\caption{(Color online) Panel (a): $A\tl{SIAM}(\w)$ for a semi-elliptic density of states, half filling 
and $U=2D$ (\appref{secSIAM}). Quantities are shown in units of the full many-body bandwidth $W$.
The superposition of three Lorentz peaks $A\tl{Lor}(\w)$ has been fitted to $A\tl{SIAM}(\w)$.
Panel (b):  Corresponding Chebyshev moments. 
The result presented here was obtained with a $L=40$ fermionic chain
and CheMPS. It agrees with
the result of \ct{raas04}, see \appref{secSIAM}. 
The legend in panel (a) is valid also for panel (b).
}
\label{figChebMomLo}
\end{figure}

\figref{figChebMomLo}(a) shows the spectral density for a SIAM together with a fitted
superposition of three Lorentzians. Their difference corresponds to a background density
that is composed of either slowly varying features or features with negligible weight.
\figref{figChebMomLo}(b) shows the corresponding Chebyshev moments. The slowly 
varying background density only contributes for the first 200 moments. After that,
the Chebyshev moments for the superposition of Lorentzians starts to be a very
good approximation to the original moments,
and it seems unnecessary to compute more than
about 400 moments. For $200<n<400$, one can simply fit the analytical
expression \eqref{eqMuLo} to the original data. Using the analytical expression
with the fitted parameters, one can then continue the
 Chebyshev moments to infinity.

Fitting \eqref{eqMuLo} to the data between iterations 200 and 400 is a nonlinear
optimization problem, which can easily be solved numerically. Still, 
there exists a linear reformulation of this optimization problem, coined under the name
\tit{linear prediction} \cp{numrec07}. The linear problem 
can be analytically reformulated as a matrix inversion problem. Its solution is faster and more stable than that of the original non-linear problem. This allows in principle to  
optimize a superposition of many more Lorentzians than in the non-linear case.

\subsubsection{Linear prediction}
\label{secLinPred}

In the context of time evolution \tit{linear prediction}
has been long established in the DMRG community,\cp{white08,barthel09}
but it has only recently been applied to the computation of Chebyshev moments.\cp{ganahl14}
The optimization problem for the sequence $\mu_n$ becomes linear,
if the sequence can be defined \tit{recursively}
\eq{
\tilde{\mu}_n = - \sum_{i=1}^p a_i \mu_{n-i},   \label{eqLinPredAnsatz}
}
which is easily found to be equivalent to \eqref{eqMuLo}\cp{barthel09}.
The strategy is then as follows. Compute $n=N_c$ Chebyshev moments, and predict 
moments for higher values of $n$ using \eqref{eqLinPredAnsatz}.
The coefficients $a_i$ are optimized by minimizing the least-square error
$\sum_{n\in \mcal N_\tx{fit} } |\tilde \mu_n-\mu_n|^2$
for a subset $\mcal N_\tx{fit}=\{N_c-n\tl{fit},\dots,N_c-1,N_c\}$ of the computed data.
We confirmed $n\tl{fit}=N_c/2$ to be a robust choice,\cp{barthel09,ganahl14} small enough
to go beyond spurious short-time behavior and
large enough to have a good statistics for the fit. Minimization yields
\eq{
R \bm{a} & = -\bm{r}, \quad \bm{a}=-R^{-1}\bm{r}, \\ 
R_{ji} & = \sum_{n\in\mcal N_\tx{fit}} \mu^*_{n-j} \mu_{n-i},\quad
r_{j} = \sum_{n\in\mcal N_\tx{fit}} \mu^*_{n-j} \mu_n. \non
}
We found that linear prediction loses its favorable filter
properties if choosing $p$ to be very high. Therefore one should
restrict the number of Lorentzians to $p=\min(n\tl{fit}/2,100)$.
Furthermore, one adds a small constant $\delta=10^{-6}$ to 
the diagonal of $R$ in order to enable the inversion of the 
singular matrix $R$.
Defining\cp{barthel09}
\[M=\left(\begin{array}{ccccc}
-a_1 & -a_2 & -a_3 & \dots & -a_{p}\\
1   & 0     &   0  & \dots & 0\\
0   & 1     &   0  & \dots & 0\\
\vdots   & \ddots  & \ddots &\ddots & \vdots\\
0  & 0       & \dots &1 & 0\\
\end{array}\right),
\]
one obtains the predicted moments 
$\tilde\mu_{N_c+n} = (M^n \bm{\mu}_{N_c})$, where 
$\bm{\mu}_{N_c} = (\mu_{N_c-1}\; \mu_{N_c-2}\; \dots\; \mu_{N_c-p})^T$. 
The matrix $M$ usually has eigenvalues with absolute value larger than $1$,
either due to numerical inaccuracies or due to the fact that linear prediction
cannot be applied as $\mu_n$ rather increases than decreases on the 
training subset $\mcal N_\tx{fit}$.
In order to obtain a convergent prediction, we set the weights 
that correspond to these eigenvalues to zero measuring the ratio of
the associated discarded weight compared to the 
total weight. If this ratio is higher than a few percent, 
we conclude that linear prediction cannot yet be applied and restart the Chebyshev 
calculation to increase the number of computed moments $N_c$.

\subsubsection{Failure of linear prediction}

It is not \tit{a priori} clear that
the spectral function can be well approximated by a superposition of Lorentzians,
although this is true for the SIAM as shown in \figref{figChebMomLo}. Other 
types of smooth functions lead to a different functional dependence 
of the moments on $n$ than the exponentially damped behavior. Close to 
phase transitions, \eg one might find an algebraic decay in the time evolution,
corresponding to an algebraic decay in the Chebyshev moments. 
If the spectral function has rather Gaussian shaped peaks, the decrease
of Chebyshev moments is $\propto e^{-(\s n)^2}$ (\appref{secMuLorentzGauss}).
For both scenarios, linear prediction is a non-controlled extrapolation scheme.
It still extracts oscillation frequencies (peak positions)  with high reliability, but predicts
a wrong decrease of the envelope, which often leads to an overestimation of peak weights. 

In practice it turns out that a combination of \tit{damping} 
with a Jackson kernel (Kernel Polynomial Method) 
and \tit{linear prediction} is a powerful way to get controlled estimates for 
the spectral function. While damping always underestimates peak heights, 
linear prediction typically overestimates peak heights.  
Both methods trivially converge to the exact result, when $N_c\ra\infty$. 
One therefore obtains upper and lower bounds 
for the spectral function. This is particularly valuable
in the DMFT as overestimated (diverging) peak heights can spoil convergence of 
the DMFT loop.

A historically much used alternative to linear prediction, suitable for arbitrary forms of the 
spectral function, is an extrapolation of Chebyshev moments 
using maximum entropy methods \cp{silver97}. 
These suffer from severe numerical instabilities, though. Of course, one might
also think of fitting another ansatz than the one of the exponential decrease.
As it is \tit{a priori} not clear which ansatz should be better, it is meaningful to stick
to the easily implemented linear prediction that is moreover known to be 
applicable for the description of quasi-particle features.

\section{Results for DMFT calculations with two baths}
\label{secVBDMFT}
\subsection{Results for two-site DCA (VBDMFT)}
\label{secVBDMFT}

In order to benchmark the Chebyshev technique
for a two-bath situation, which goes beyond previous work\cp{ganahl14} 
(see \appref{secOneBath}), we study the
Hubbard model on the two dimensional square lattice
\eq{ \label{eqHubbard}
H\tl{Hub} & = \sum_{\bm k\s} \varepsilon_{\bm k} c_{\bm k,\s}\dag c_{\bm k,\s} + U \sum_i n_{i\uparrow} n_{i\downarrow}, \\
 \varepsilon_{\bm k} & =  - 2 t (\cos(\bm{k}_x) + \cos(\bm{k}_y)) - 4t'\cos(\bm{k}_x) \cos(\bm{k}_y).  \non
}
in a two-site dynamical cluster approximation\cp{maier05} (DCA) developed by 
\ct{ferrero09}. This so-called \tit{valence bond} DMFT (VBDMFT)
is a minimal description of the normal phase of the
high-temperature superconductors, using a minimal two patches DCA cluster.
It leads to a simple physical picture of the pseudogap phase in terms of a selective Mott transition
in the momentum space.
We choose this model here as a benchmark since its solution
contains low energy features in the spectral functions (pseudogap),  
which have required high-precision QMC computations
followed by a careful Pad\'e analytic continuation.
Moreover, real-frequency computations are very important for the comparison with experiments
that measure e.g. the optical conductivity along c-axis.\cp{ferrero10}  
It is therefore a non-trivial case where DMRG 
impurity solvers would bring significant improvements
over the QMC in practice.

To set up the VBDMFT, one splits the Brioullin zone
into a \tit{central} patch $P_+ = \{\bm{k}\, \big | \abs{k_x} < k_0 \wedge 
\abs{k_y} < k_0\}$, where $k_0=\pi(1-1/\sqrt{2})$, 
and a \tit{border} patch $P_{-}=\{\bm{k}\, \big | \bm{k} \notin P_{+} \}$. 
In the DCA, the $\bm{k}$-dependence of the self-energy $\Sigma_\k(\w)$ 
within each patch is neglected and one computes a Green's function
for a patch by averaging over all $\bm{k}$ vectors in the patch
\eqs{ \label{eqGpatch}
 G_{\k}(\w) & = \frac{1}{\abs{P_\k}} \sum_{\bm k \in P_\k} \frac{1}{\w + \mu - \varepsilon_{\bm k} - \Sigma_\k(\w)}, \\
 \Sigma_\k(\w) & = G_{0\k}(\w)^{-1} - G_{\k}(\w)^{-1}.
} 

Representing the non-interacting baths in a chain-geometry,
and taking the two impurities to be the first of 
two chains $c_{\kappa\sigma} \equiv c_{0\kappa\sigma}$,
the model Hamiltonian that needs to be solved is
\eq{ 
   H  & = H_d + H_{b,+} + H_{b,-}   \non\\
  H_d & =
  \sum_{\kappa = \pm \atop \sigma = \uparrow,\downarrow} 
  \left(
  \ol t_\k +\varepsilon_0 
  \right)
  n_{\k\s}   
  + 
  \frac{U}{2}  \sum_{\kappa = \pm \atop \ol \kappa = -\kappa}  \big(
  n_{\kappa\uparrow} n_{\kappa\downarrow}  +  n_{ \kappa \uparrow} n_{\ol \kappa\downarrow}   \non\\
  & \qquad +  c^{\dagger}_{\kappa\uparrow} c^{\dagger}_{\kappa\downarrow}  c_{\ol \kappa\downarrow} c_{\ol \kappa \uparrow}
  + c^{\dagger}_{\kappa\uparrow} c^{\dagger}_{\ol \kappa\downarrow}  c^{}_{\kappa\downarrow} c^{}_{\ol \kappa\uparrow}
  \big),   \label{eqHvbdmft} \\
  H_{b,\k} & = \sum_{i=0,\sigma}^{L_\k} 
   	    t_{i\k} ( c_{i\k\sigma}\dag c_{i+1,\k\sigma} + \tx{h.c.} ) + \sum_{i=1,\sigma}^{L_\k}  \varepsilon_{i\k} n_{i\k\sigma} ,  \non 
}
where $\varepsilon_0=-\mu$ and the term $\ol t_\k =  \frac{1}{\abs{P_\k}}  \sum_{k\in P_\k} \varepsilon_k$ 
accounts for high-frequency contributions of the hybridization function (see \appref{secHVBDMFT}). 

The $\k$-space interaction term in \eqref{eqHvbdmft} arises when diagonalizing the
hybridization function of a real-space two-site cluster
$c_{\pm\s} = \frac{1}{\sqrt{2}} (c_{1\s} \pm c_{2\s})$, where $c_{1\s}, c_{2\s}$ are annihilation
operators for the cluster sites in real-space, and
$c_{\pm\s}$ for the cluster sites in $\k$ space. 
In real-space, the interaction is a simple Hubbard expression, 
but then the hybridization function is non-diagonal. A diagonal hybridization function, which 
leads to two uncoupled baths for the patches and by that allows a simple  
chain-geometry for the whole system, is therefore only possible in $\k$-space.
The more complex form of the interaction in $\k$-space does
not affect the efficiency of DMRG.

We iteratively solve the self-consistency equation 
obtained by inserting the self-energy estimates 
of the impurity model \eqref{eqHvbdmft} 
into the lattice Green functions \eqref{eqGpatch}.
We do that on the real-energy axis 
with an unbiased energy resolution. The details of this calculation are described in   
\appref{secVBDMFTdet}.

\begin{figure}
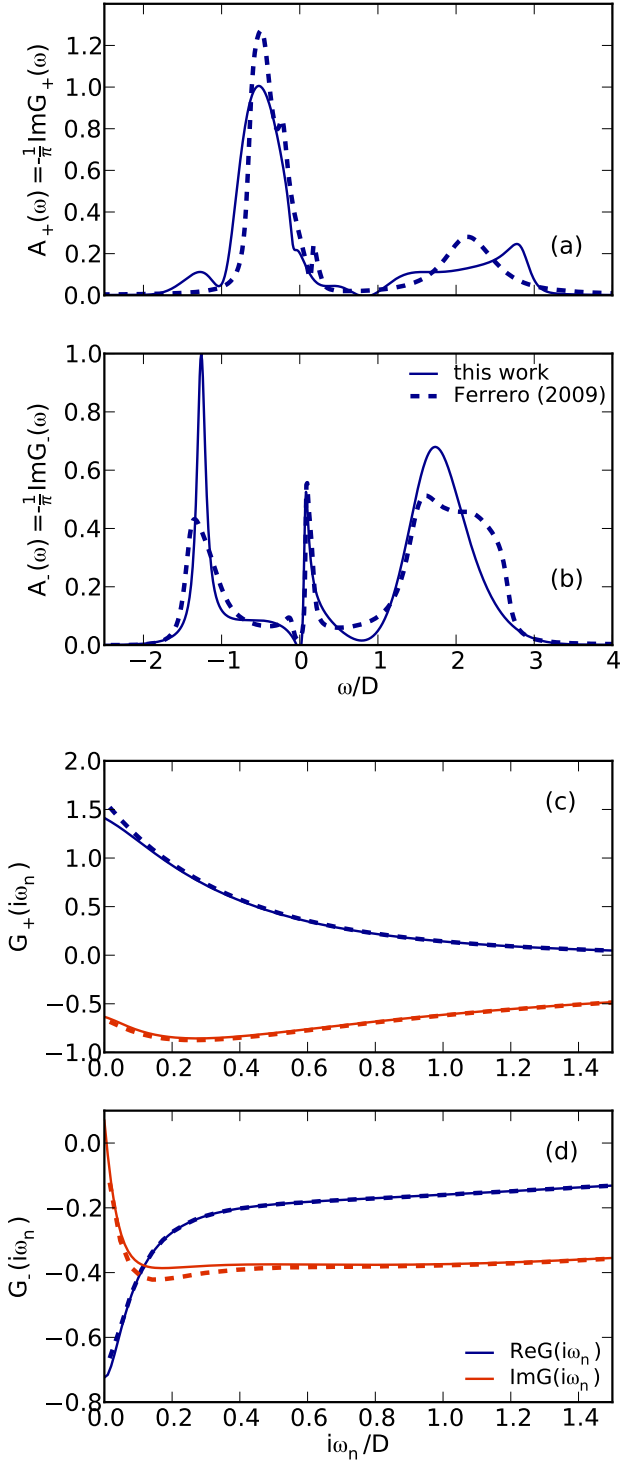

\ig{0.48}{pdf}{vbdmft_rho}
\ig{0.48}{pdf}{vbdmft_mat}
\caption{(Color online) Spectral functions (a,b) and Green's functions on the imaginary axis (c,d)  
within VBDMFT \cp{ferrero09} for $U=2.5D$ and $n=0.96$.
We compare our zero-temperature CheMPS results (solid lines) with CTQMC data for $T=1/200$ (dashed lines) from \ct{ferrero09}. For this computation, we used the $b=0$ setup, a chain length of $L=30$ per bath, a 
truncation error of $\ve\tl{compr}=10^{-3}$, and $N/a=60/D$, $a=40D$.}
\label{figVBDMFTrho}
\end{figure}

In \figref{figVBDMFTrho}(a) and (b), we compare our CheMPS results for the spectral densities of the two 
momentum patches with those of \ct{ferrero09} obtained using 
CTQMC and analytical continuation. 
We observe a good overall agreement between the two methods, in particular at low frequencies.
Low energy features (pseudogap), in particular in $A_{-}(\w)$, are well reproduced by both methods.
At high energy (Hubbard bands) however, there are some  differences between QMC and CheMPS
(and also between the two variants of CheMPS). This is to be expected since
the Pad\'e analytic continuation technique used on the QMC data in Ref.~\onlinecite{ferrero09} 
is not a precision method at high energy.
\begin{figure}
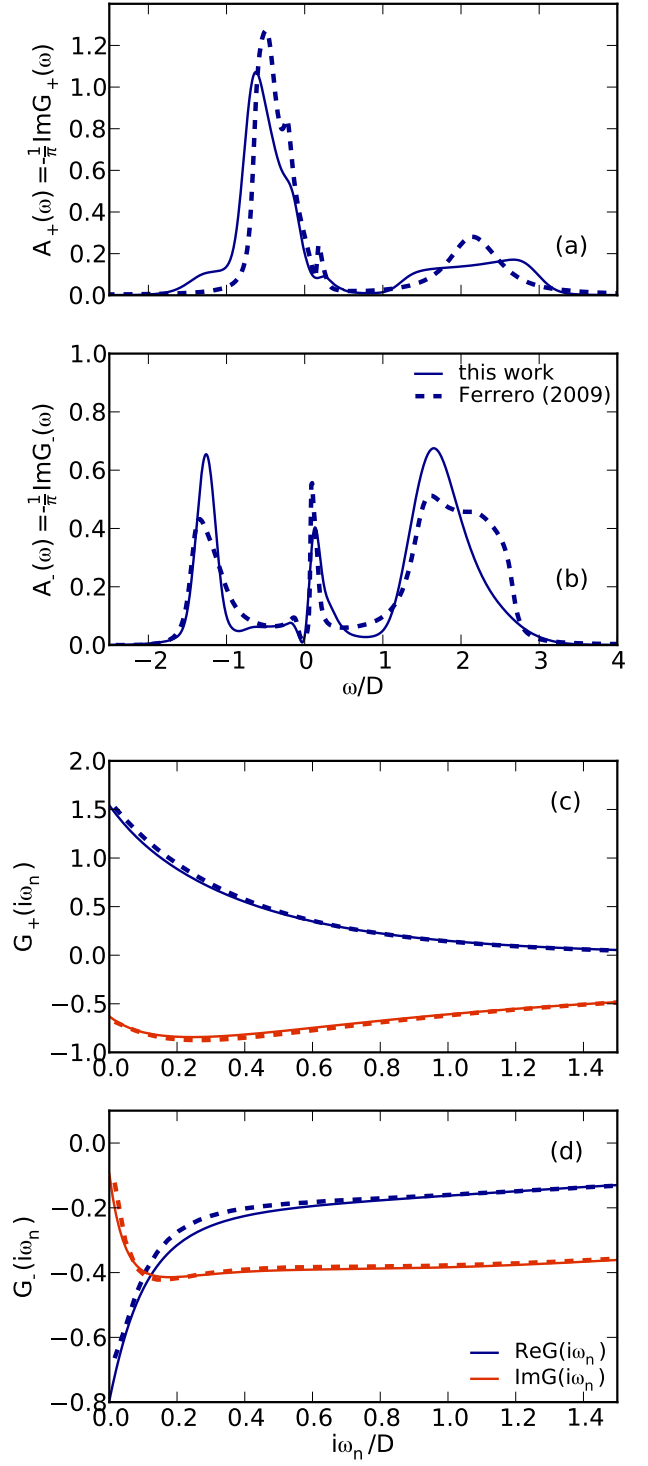

\ig{0.48}{pdf}{vbdmft_rhob-a}
\ig{0.48}{pdf}{vbdmft_matb-a}
\caption{(Color online) The same comparison as in \figref{figVBDMFTrho}. 
For this computation, we used the $b=-a$ setup, a chain length of $L=40$ per bath, a 
truncation error of $\ve\tl{compr}=10^{-3}$, and $N=450$ and $a=15D$. 
For the $b=-a$ setup, one can use a smaller value of $a$ as 
in the $b=0$ setup, as discussed in \secref{secOptCheb}.}
\label{figVBDMFTrhob}
\end{figure}

In \figref{figVBDMFTrho}(c) and (d), we do the analogous comparison on the imaginary axis,
and find much better agreement.
On the imaginary axis, the QMC results can be considered numerically
exact. The very low temperature ($\beta D=200$) used for QMC should yield results that are 
indistinguishable from a zero-temperature calculation. 
The slight disagreement of our data and the QMC data on the Matsubara
axis could probably be removed if we were able to reach higher expansion orders. 
One DMFT iteration for the presented $b=0$ calculation 
took around \SI{5}{h} running on four cores with \SI{2.5}{GHz}. Convergence 
is achieved after 10 iterations starting from the non-interacting solution. 
The calculation has been carried out with two attached chains of $L=30$ lattice sites each. 
We did not observe changes for higher bath sizes up to $L=40$, but could not reach 
high enough expansion orders for chains longer than $L=40$. 
We computed $N=2500$ moments using a scaling constant $a=40D$, 
which corresponds to the full bandwidth.

The calculation can be accelerated significantly by using the $b=-a$ 
setup of \secref{secba} and avoiding \tit{linear prediction}. 
This leads to the same quality of agreement with QMC 
on the Matsubara axis, but on the real axis, peaks are a bit less
pronounced while the pseudogap is still well resolved (\figref{figVBDMFTrhob}). 
While the study of systems with higher bath sizes increases 
the computational cost tremendously in the $b=0$ setup,
we could easily go to $L=50$ within the $b=-a$ setup. 
This did not change the results. Computation times 
varied from \SI{1.2}{h} per iteration for $L=30$, 
over  \SI{3}{h} for $L=40$ to around \SI{10}{h} for the $L=50$ calculation.
We computed $N=450$ moments using a scaling of $a=15D$ in all cases.

\subsection{Single-site two-orbital DMFT}
\label{secTIAM}

In the following, we apply CheMPS to the DMFT treatment of the 
two-orbital Hubbard model
\eq{
H
     & =
     \sum_{\bm{k}\nu\sigma}
     \ve_{\bm{k}\nu}
     n_{\bm{k}\nu\sigma}
     +
     U\sum_{i\nu}n_{i\nu\uparrow}n_{i\nu\downarrow} \non\\
      & +
     \sum_{i\sigma\sigma'}
     (U_{1}-\delta_{\sigma\sigma'}J)
     \,
      n_{i1\sigma}n_{i2\sigma'}
     \nonumber \\
     & + \frac{J}{2}
     \sum_{i\nu\sigma}
     c_{i\nu\sigma}^{\dag}
     (
     c_{i\ol{\nu}\,\ol{\sigma}}\dag c_{i\nu\ol{\sigma}}
     +
     c_{i\nu\ol{\sigma}}\dag c_{i\ol{\nu}\,\ol{\sigma}}
     )
     c_{i\ol{\nu}\sigma}\dag
}
on the Bethe lattice. We study a parameter regime 
close to the Metal-Insulator phase transition. 
This regime is computationally particularly
expensive and we had to use a 
logarithmic discretization to reach Chebyshev expansion 
orders at which spectral functions 
are completely converged with respect to 
expansion order and system size. 
The linear discretization was feasible in
the case of the VBDMFT studied in the previous section,
as there, we faced a smaller entanglement entropy production
during Chebyshev iterations. 
\begin{figure}
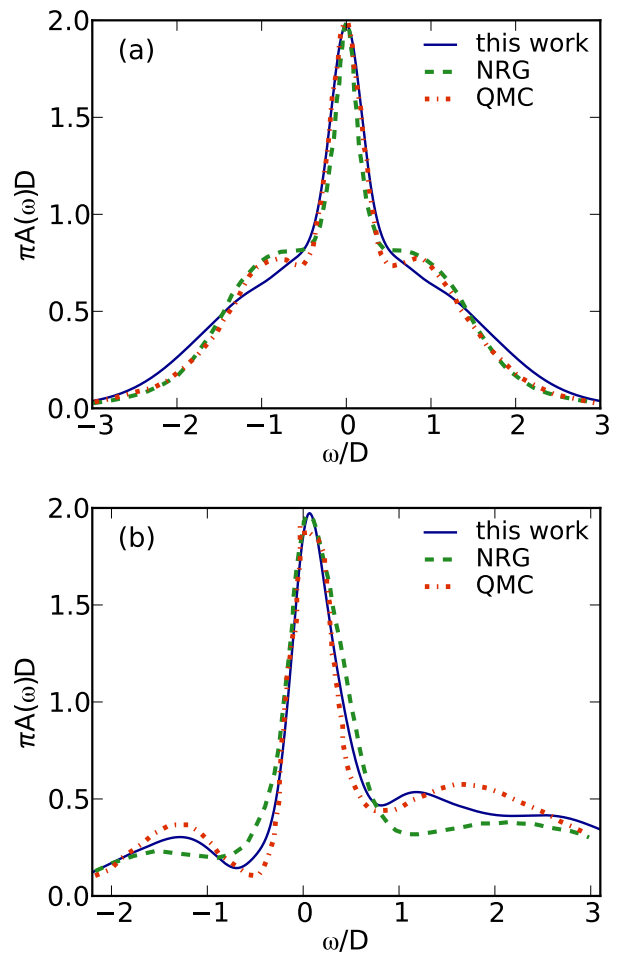

\ig{0.48}{pdf}{stadler_half}
\ig{0.48}{pdf}{stadler_dope}
\caption{(Color online) Spectral function for the two-band Hubbard model.
Panel (a): $U/D=1.6$, $n = 2$ (half filling). Panel (b) $U/D=3.8$, $n=1$ (quarter filling). 
In both cases $J=\frac{1}{6}U$, $U'=U-2J$. 
We fixed a truncation error $\ve\tl{compr} = 10^{-3}$, used
a scaling $a=25D$, computed $N_c=150$ moments and used linear prediction. 
To represent the two baths, 
we used two chains of length $L=20$ each, obtained with a  
logarithmic discretization parameter
of $\Lambda=2$, leading to 
grid energies $\Lambda^{-n}$ (see \eg Ref.\ \onlinecite{bulla08}).
The NRG calculation was done for temperature $T/D=0.0025$, the QMC calculation for $T/D=0.01$. 
Both should be almost indistinguishable from a $T=0$ calculation. 
NRG data from K. Stadler\cp{stadler13} computed with a code of A. Weichselbaum.\cp{weichselbaum12}
 QMC data from M. Ferrero.\cp{ferrero14}
}
\label{figTIAMstad}
\end{figure}

Using a logarithmic discretization is not necessary for CheMPS.
But as it leads to exponentially decaying hopping constants, it gives rise to three  
advantages: 
(i) One can use smaller scaling constants $a$ 
as the many-body bandwidth is considerably reduced due to the exponentially
small value of most hopping constants in the system.
(ii) One faces a smaller entanglement entropy production: 
at the edges of the bath chains (far away from the impurity),  
hopping constants are exponentially small, and application of $H$ therefore 
creates much less entanglement than in the case in which a linear discretization
is used. In \eqref{eqtRec}, the action of $H'$ on $\ket{t_{n-1}}$ is then only a
small perturbation for most parts of the system, and the recursion is therefore
dominated by the second term $\ket{t_{n-2}}$. Entanglement therefore builds up
only in the region where it is relevant, that is, in the vicinity
of the impurity. Hence, matrix dimensions grow considerably more slowly when using 
a logarithmic discretization as compared to a linear discretization. 
(iii) One faces a faster speed of convergence of the Chebyshev moments as in the linear case:  
The complexity of the spectral function is considerably reduced when 
averaging over possible peaks in the high-energy structure of the spectral function, as is 
done when using a logarithmic grid. The associated Chebyshev expansion therefore
converges more quickly than in the case of a linear grid.

When using a logarithmic discretization, one has to convolute 
the resulting spectral function with a Gaussian\cp{weichselbaum07} 
to average over the finite-size features 
that originate from the coarse log resolution at high energies.
\begin{figure}
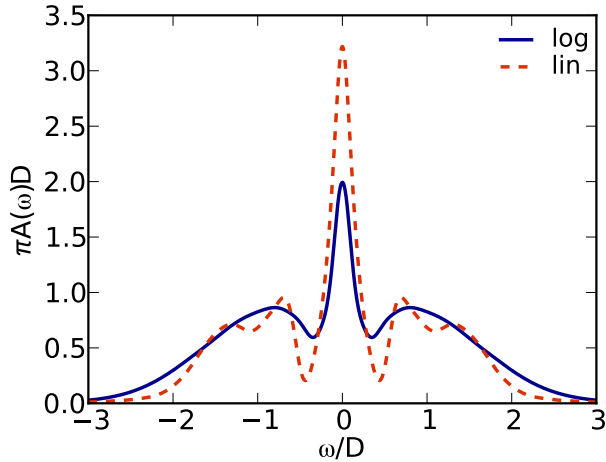

\ig{0.48}{pdf}{tiam_gan}
\caption{(Color online) Spectral function for the two-band Hubbard model.
The system parameters $U/D=1.6$, $J/U=\frac{1}{4}$, $U'=U-2J$, $n=2$ are very similar
to the one in \figref{figTIAMstad}(a). 
We performed a calculation with linear (``lin", $L=40$ per bath) 
and one with logarithmic discretization (``log", $L=20$ per bath).
We fixed a truncation error $\ve\tl{compr} = 10^{-3}$. 
For the calculation with logarithmic discretization, we used
a scaling $a=25D$ and computed $N_c=300$ moments. 
For the calculation with linear discretization, we used 
a scaling of $a=125D$ and computed $N_c=1250$ moments. 
We used linear prediction in all cases.
The logarithmic discretization used a discretization parameter
$\Lambda=2$, leading to grid energies $\Lambda^{-n}$ (see \eg Ref.\ \onlinecite{bulla08}).
}
\label{figTIAMgan}
\end{figure}

In \figref{figTIAMstad}, 
we compare exemplary calculations
for the two-band Hubbard model with NRG and analytically continued QMC data. 
We find good agreement in the regions around the Fermi energy,
where the pinning criterion is respected to high accuracy without being enforced. 
We explain the observed disagreement far away from the Fermi energy
with a different specific implementation of the broadening convolution. 
One DMFT iteration for our
calculations took around \SI{20}{min} running on two \SI{2.5}{GHz} cores.

In \figref{figTIAMgan}, we study the case of Ref.~\onlinecite{ganahl14i}, which
is very similar to the one studied in \figref{figTIAMstad}(a). 
Our results suggest that the data shown in Ref.~\onlinecite{ganahl14i}
is not fully converged with respect to computed time in tDMRG,
as it does not fulfill the pinning criterion. 
We face a similar problem when using a linear discretization:  
For the reachable Chebyshev expansion orders,
we do not observe convergence of the central peak height 
for increasing expansion orders. All peaks, side peaks as well
as central peak, increase for increasing 
expansion order and the pinning criterion is not fulfilled. 
The additional structure in the Hubbard band,
which is not visible in the calculation with the logarithmic discretization,
is seen to be similar to the one observed in Ref.~\onlinecite{ganahl14i}.
One DMFT iteration for the computation that uses a logarithmic grid took \SI{20}{min}
running on two \SI{2.5}{GHz} cores. For the linear grid this time was \SI{10}{h} per 
DMFT iteration.
\begin{figure}
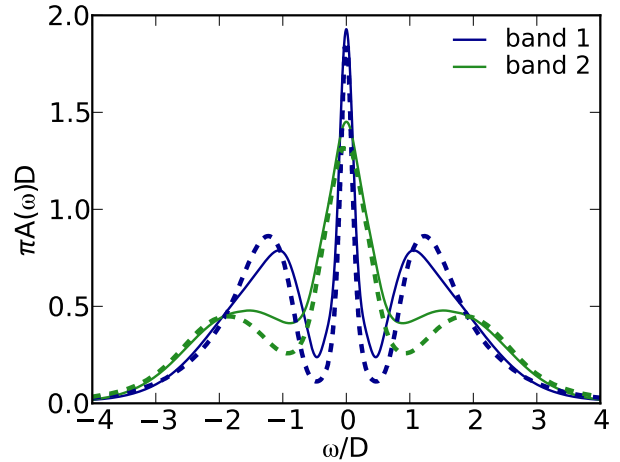

\ig{0.48}{pdf}{tiam}
\caption{(Color online) 
Results for the spectral densities in the two orbitals for $J=0$. Our results are 
for $U=2.6D$, $U'=1.3D$, $n=2$ (half filling) and depicted by the solid lines. 
The reference NRG results\cp{greger13} are for 
$U=2.8D$, $U'=1.4D$ and depicted by the dashed lines. We had to choose a slightly smaller interaction for 
a meaningful comparison, as for the
parameters of \ct{greger13}, we converged, though very slowly, into
an insulating solution without central peak.
The non-interacting single-particle half-bandwidth of the first band 
is $D$, and the one of the second band is $1.4D$.
We used two chains of length $L=20$ each, and a logarithmic discretization parameter
of $\Lambda=2$, leading to grid energies $\Lambda^{-n}$ (see \eg Ref.\ \onlinecite{bulla08}).}
\label{figTIAM}
\end{figure}

Finally, we study parameters that lead to a system close to the 
metal-insulator phase transition. \figref{figTIAM} shows that 
we obtain satisfactory agreement with NRG data, given the fact
that we had to reduce the interaction slightly in order to stay in the metallic phase. 
This slight quantitative mismatch can possibly again be explained 
with a differing broadening convolutions in the two calculations. 
One DMFT iteration took \SI{2}{h} for the calculation
of \figref{figTIAM}, when fixing a truncated weight 
of $\ve\tl{compr}=2\times 10^{-3}$.

\section{Conclusions}
\label{secConc}

We solved several DMFT problems with two baths 
on the real frequency axis with unbiased energy-resolution based on an DMRG impurity solver using 
Chebyshev polynomials for the representation of spectral functions at moderate numerical effort. DMRG is thereby seen to be a viable alternative for DMFT impurity solvers also beyond the well-understood single-impurity single-band case. 

Technically, it was crucial to apply the adaptive truncation 
scheme of \secref{secCheMPS} to maintain a modest numerical effort: 
in all cases, the new scheme gave much better results than the previously 
employed scheme based on fixed matrix dimensions. 
Another important way of tuning the calculation is provided 
by the mapping of the spectrum to the convergence interval of Chebyshev polynomials: The 
different options to set up a CheMPS calculation can be summarized
to yield two alternatives. (i) One uses the $b=0$ setup
and post-processes moments with linear prediction. (ii) One uses the $b=-a$ setup and 
avoids linear prediction, using simple Jackson damping. Depending on the problem,
the first or the second method can be more efficient. The second alternative 
is computationally much more efficient for cases in which 
linear prediction is a non-controlled extrapolation scheme,
but has problems to resolve sharp peaks at the Fermi edge.

The method presented in this paper can in principle be extended to 
the case of more than two baths without major changes to the DMFT-DMRG 
interface and the Chebyshev-based impurity solver as such. 
However, while two baths can still be modeled by a single chain 
with the impurity at the center (instead of at the end, as in single-band DMFT), 
this is no longer possible for three and more baths. 
This will necessitate a new setup of the DMRG calculation replacing 
the chain-like by a star-like geometry with the impurity at the center of the star, 
hence a generalization from a matrix-based to a tensor-based 
representation at the location of the impurity. It remains to be seen at 
which numerical cost reliable results on the real frequency axis will be obtainable.

\section{Acknowledgements}

FAW acknowledges discussions with C. Hubig and K. Stadler. 
US acknowledges discussions with M. Ganahl and H.-G. Evertz. 
FAW and US acknowledge discussions with P. Werner and 
support by the research unit FOR 1807 of the DFG.
O. P. acknowledges support from the ERC Starting Grant 278472--MottMetals.
We acknowledge K. Stadler and M. Ferrero for providing the data 
of their NRG and QMC calculations.

\appendix

\section{Scaling of Chebyshev Moments with respect to energy scaling}
\label{secScaling}

The Chebyshev moments obtained by using
two different scalings $H_1' = H/a_1$ and $H_2'=H/a_2$ are
from \eqref{eqMuH} $\mu_n^{a_1}=\sum_i W_i T_n((E_i-E_0)/a_1)$ 
and $\mu_n^{a_2}=\sum_i W_i T_n((E_i-E_0)/a_2)$.
As we consider one-particle operators $c\dag$ 
the weights $W_i=\vt\bra{E_i} c\dag \ket{E_0}\vt^2$ fulfill 
\eq{
W_i = 0 \tx{ for } E_i 
\tx{ with } \abs{E_i-E_0} \gtrsim W\tl{single},
}
where $W\tl{single}$ is the single-particle bandwidth.
If the scalings  $a=\min(a_1,a_2)$ are chosen large enough, $W\tl{single}/a \ll 1$, then 
\eq{ \label{eqScaling}
\mu_{a_1n}^{1} = \mu_{a_2n}^{2} \quad\tx{ if } \frac{a_1n}{4} \in \mathbb{N} \tx{ and } \frac{a_2n}{4} \in \mathbb{N}.
}

Proof: If these requirements are met, the eigenvalues $E_i$ with $W_i \neq 0$ 
are close to the groundstate energy: $x=(E_i-E_0)/a \ll 1$. 
The Taylor expansion $\arccos(x) = \pi/2 - x - x^3/6 + \dots$ becomes reliable already
when $x\lesssim \frac{1}{2}$, which is fulfilled if $a$ is at least 
twice the single-particle bandwidth as in all hitherto known applications\cp{weisse06,holzner11,ganahl14}.

Consider a particular energy $E=E_i-E_0$ for which $W_i>0$. It holds
\eqn{ 
T_{a_1 n}(E/a_1) & = T_{a_2 n}(E/a_2) \\ 
\cos(a_1 n \arccos(E/a_1)) &= \cos(a_2 n \arccos(E/a_2)) \\
\cos(a_1 n (\pi/2 - E/a_1)) & \simeq \cos(a_2 n (\pi/2 - E/a_2)) \\
a_1 n (\pi/2 - E/a_1) \tx{ mod } 2\pi & \simeq a_2 n (\pi/2 - E/a_2) \tx{ mod } 2\pi \\
a_1 n \pi/2  \tx{ mod } 2\pi & \simeq a_2 n \pi/2 \tx{ mod } 2\pi \\
a_1 n /2  \tx{ mod } 2 & \simeq a_2 n /2 \tx{ mod } 2.
}
A sufficient condition for the last line to hold is that both $a_1 n /2$ and $a_2 n /2$ are multiples
of 2, \ie the statement of \eqref{eqScaling}.

\section{Chebyshev Moments of Lorentzian and Gaussian}
\label{secMuLorentzGauss}

If we fix the shift to be $b=0$, 
equation \eqref{eqMuLo} is obtained as follows. 
As $\mu_n = \sum_i \alpha_i \mu_n^{li}$ we only have 
to compute the moments for a single Lorentzian, which allows to drop the index $i$
\eqn{
\mu_n^l 
& = \frac{\eta}{\pi} \int_{-1}^{1} d\w \frac{\cos(n\arccos(\w))}{(\w-\w_0)^2+\eta^2} \\
& \simeq \frac{\eta}{\pi} \int_{-1}^{1} d\w \frac{\cos(n(\frac{\pi}{2}-\w))}{(\w-\w_0)^2+\eta^2} \\
& = \frac{\eta}{\pi} \int_{-1}^{1} d\w \frac{\cos(n(\w+\w_0'))}{\w^2+\eta^2}, ~~ \w_0' = \w_0-\frac{\pi}{2} \\
& = \frac{\eta}{\pi} \tx{Re} \int_{-1}^{1} d\w \frac{\exp(in(\w+\w_0'))}{\w^2+\eta^2}\\
& = \frac{\eta}{\pi} 2\pi i\, \tx{Res} \Big( \frac{\cos(in(\w+\w_0'))}{\w^2+\eta^2} \Big)\Big\vert_{\w=i\eta}\\
& = \cos(n(\w_0-\frac{\pi}{2})) e^{-n\eta}.
}
When closing the integral in the complex plane, we assumed that the Lorentzian concentrates
almost all of its weight within $[-1,1]$, which is a meaningful assumption, as we are calculating
with the rescaled frequencies.

For the Gaussian one has
\eq{
  A\tl{Gauss}(\w) & = \sum_i \alpha_i \frac{1}{\sqrt{2\pi}\s_i} e^{-\frac{(\w-\w_i)^2}{2\s_i^2}}, \non\\
 \Ra \quad \mu_n^g   & \simeq \sum_i \alpha_i \cos(n(\w_i-\frac{\pi}{2})) e^{-(\s_in)^2/2},  \label{eqMuGauss}
}
as shown by a similar calculation:
\eqn{
\mu_n^g
& = \frac{1}{\sqrt{2\pi}\s} \int_{-1}^{1} d\w\,  e^{-\frac{(\w-\w_0)^2}{2\s^2}} \cos(n\arccos(\w)) \\
& = \frac{1}{\sqrt{2\pi}\s} \int_{-1}^{1} d\w\,  e^{-\frac{\w^2}{2\s^2}} \cos(n(\w+\w_0')), ~ \w_0' = \w_0-\frac{\pi}{2} \\
& = \frac{1}{\sqrt{2\pi}\s} \tx{Re} \int_{-1}^{1} d\w\,  e^{-\frac{\w^2}{2\s^2} + in \w+ in\w_0'} \\
& = \tx{Re} \, e^{-\frac{\sigma^2 n^2}{2} + i n \w_0'} = \cos(n(\w_0-\frac{\pi}{2})) e^{-\frac{\sigma^2 n^2}{2}}.
}
From the third to the fourth line, the extension of the integral limits to $\pm\infty$ in order to apply
the Gaussian integral formula is well justified, as the Gaussian concentrates all its weight within $[-1,1]$.

\section{Single-bath impurity calculations}
\label{secOneBath}

\subsection{Single-impurity Anderson Model}
\label{secSIAM}

The single impurity Anderson model (SIAM) in its
truncated chain representation is 
\eq{
  H=&
  \sum_{n=0,\sigma}^{L-1} t_n(c_{n\sigma}^{\dagger}c_{n+1\sigma}+\tx{h.c.})
  +  \sum_{n=0,\sigma}^{L} \ve_i n_{0\sigma}
  +U n_{0\downarrow}n_{0\uparrow},    \label{eqSIAM}
}
with hybridization function\cp{raas05}
\eq{
\Delta(z) = \frac{t_0^2}{z - \ve_1 - \displaystyle\frac{t_1^2}{\displaystyle z-\ve_2-\frac{\cdots}{z-\ve_{L-1} -\frac{\displaystyle t_{L-1}}{z-\ve_{L}}}}}.
}
For an infinitely long chain, the continuous version of the SIAM is recovered. 
The bath density of states is $\Gamma(\w)=-\frac{1}{\pi} \tx{Im } \Delta(\w+i0^+)$. For 
an infinite homogeneous system with $t_i=t=D/2$, $\ve_i=0$, 
$\Gamma(\w)$ is the semielliptic density
of states at half bandwidth $D$\cp{raas05}
\eq{
\Gamma(\w) = \frac{2}{\pi D} \sqrt{1-(\w/D)^2}.
}
In the non-interacting case, also the spectral function $A(\w)$
is semielliptic.

\begin{figure}
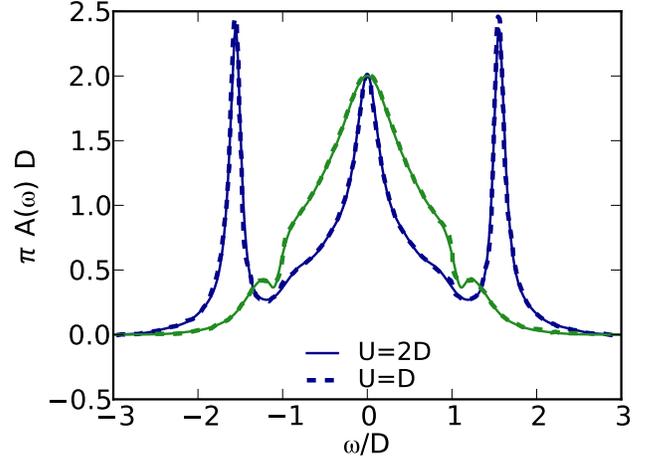

\ig{0.48}{pdf}{siam_U4}
\caption{(Color online) Single Impurity Anderson Model with semi-elliptic density of states
of half-bandwidth $D$.
We compute the spectral function with CheMPS allowing a cumulative 
truncated weight of $\ve\tl{compr}=7\times10^{-4}$
and post-process moments with linear prediction (solid lines). These results are
compared to data obtained with dynamic DMRG (dashed lines) by \ct{raas04}. 
We used a fermionic representation of the SIAM on a chain with length $L=80$.}
\label{figSIAM}
\end{figure}

The computation of the spectral function $A(\w)$ 
for the SIAM
is much less demanding than 
for most DMFT applications:  
$A(\w)$  has only few sharp features, which in addition 
are well approximated by Lorentzians (\secref{secLinPred}).
Hence, linear prediction can be applied and we observe
very good agreement with DDMRG data of \ct{raas04} in \figref{figSIAM},
confirming results of Ref.~\onlinecite{ganahl14}.
For the case $U=D$, we observe
a slight disagreement in the region of the shoulders, where
the linear prediction predicts two small peaks,
whereas DDMRG shows a perfectly flat shoulder. This
might point out a failure of linear prediction for the description of this feature.
Although this should be of minor importance here, it could matter in other cases. 

\subsection{Single-site single-orbital DMFT}
\label{secDMFT}

The single-site DMFT of the one orbital Hubbard model
\eq{
H  =
     \sum_{\bm{k}\sigma}
     \ve_{\bm{k}}
     n_{\bm{k}\sigma}
     +
     U\sum_{i\nu}n_{i\uparrow}n_{i\downarrow}
}
is well established\cp{georges96} and amounts to the determination
of the self-consistent parameters $\{t_i,\ve_i\}$ of a SIAM \eqref{eqSIAM}. 
We give a derivation of the DMFT equations only for 
the more complicated case of the cluster DMFT (\secref{secVBDMFTdet}), which
can easily be reduced to the single site case.
\begin{figure}
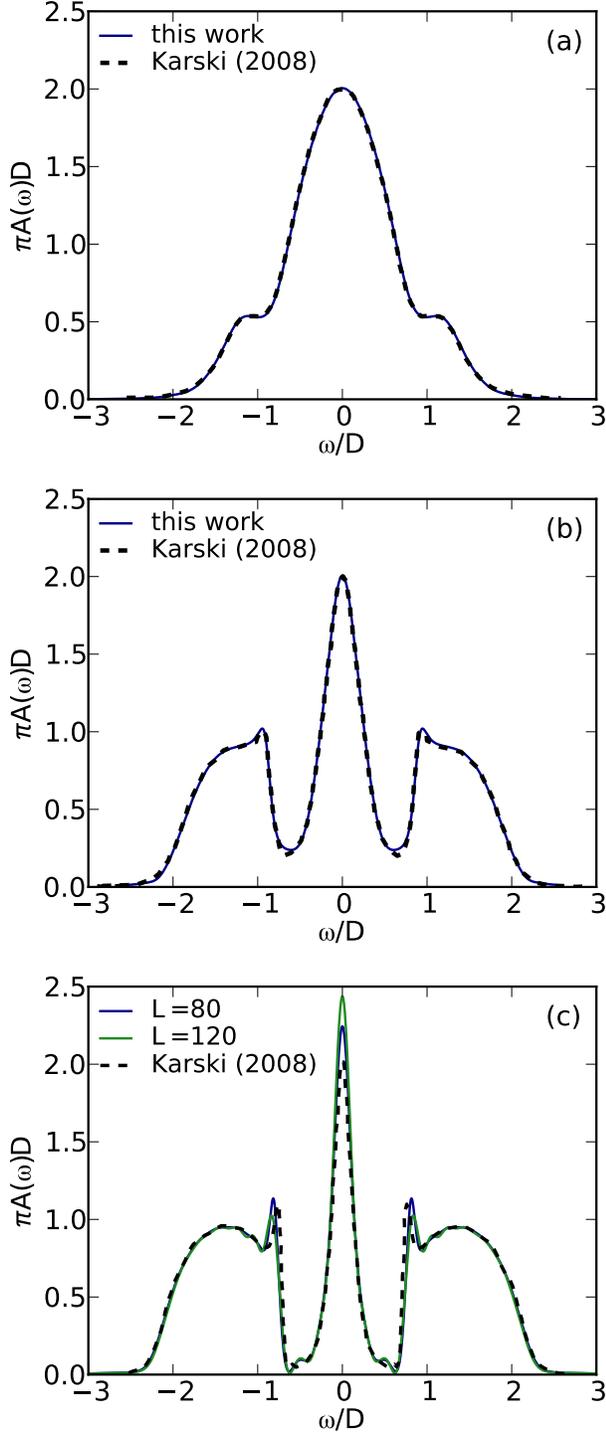

\ig{0.48}{pdf}{dmft_bethe_U2.0}
\ig{0.48}{pdf}{dmft_bethe_U4.0}
\ig{0.48}{pdf}{dmft_bethe_U4.8}
\caption{(Color online) Local density of states within DMFT 
for the single-band Hubbard model on the Bethe lattice. Computed using CheMPS with an
allowed cumulative truncated weight of $\ve\tl{compr}=5\times10^{-4}$. 
Panel (a): $U=D$. Panel (b): $U=2D$. Panel (c): $U=2.4D$. 
We compare our results with data from \ct{karski08}.}
\label{figDMFT}
\end{figure}

\figref{figDMFT} shows our results for which we fixed a maximum cumulative 
truncated weight of $\ve\tl{compr}=5\times10^{-4}$. 
For the quite featureless spectral function 
of \figref{figDMFT}(a) ($U=D$), the thermodynamic 
limit is already obtained for $L=40$ and one DMFT iteration took \SI{0.3}{h}. 
For \figref{figDMFT}(b) ($U=2D$), we needed $L=80$
and one DMFT iteration took around \SI{3}{h}. 
For \figref{figDMFT}(b) ($U=2.4D$) we obtained converged DMFT loops,
which violate the pinning criterion $A(0)=2\pi/D$, though.  
When employing large bath sizes of $L=100$ and more, we could not reach
sufficiently high numbers of Chebyshev moments within reasonable 
computation times of up to \SI{12}{h} per DMFT iteration; the linear prediction
then overestimates the height of the central peak.

\section{Technical details of VBDMFT}
\label{secVBDMFTdet}

In this appendix, we provide the technical details for the VBDMFT calculation.

\subsection{Self-consistency loop}

The Green's function for a patch $\k$ has been introduced in \secref{secVBDMFT}  and reads
\eq{ \label{eqGlatt}
 G_{\k}(z) & = \frac{1}{\abs{P_\k}} \sum_{\bm k \in P_\k} \frac{1}{z + \mu - \varepsilon_{\bm k} - \Sigma_\k(z)}.
}
Within the DCA, one obtains an estimate for $\Sigma_\k(z)$ by solving
an auxiliary impurity-bath system, the Green's function of which is 
\eq{ \label{eqGimp}
 G_{\k}\th{imp}(z)^{-1} &= z + \mu -\Delta_\k(z)-\Sigma_\k(z),
}
where the bath is completely characterized by the hybridization function $\Delta_\k(z)$.

The problem is then to determine $\Delta_\k(z)$ such that the impurity-bath 
system best approximates the actual lattice environment, which amounts to  
 the \tit{self-consistency} condition
\eq{ \label{eqSelfCons}
G_{\k}(z) =  G_{\k}\th{imp}(z).
}
This equation constitutes a fixed-point problem for the hybridization
function $\Delta(z)$ and can hence be solved iteratively, starting with some
initial guess, \eg the non-interacting solution. 

Solving the impurity problem for the initial guess of $\Delta(z)$, one obtains
$G_{\k}\th{imp}(z)$. From that one obtains the estimate for the self-energy as
$\Sigma_\k(z)  = G_{0\k}\th{imp}(z)^{-1} - G_{\k}\th{imp}(z)^{-1}$,
or by the method of \ct{bulla98} (we found the latter not to 
yield advantages for the CheMPS setup). The self-energy is then
inserted into \eqref{eqGlatt} to obtain a new value for $G_{\k}(z)$.
Using self-consistency, this  defines a new hybridization
function by inserting \eqref{eqSelfCons} in \eqref{eqGimp}:
\eq{ \label{eqNewDelta}
\Delta_\k(z) = - G_{\k}(z)^{-1} + z + \mu - \ve_0  - \Sigma_\k(z).
}

In QMC calculations, one defines all quantities on the imaginary axis. In this
work as in NRG calculations, we define all quantities
on the real axis: the spectral density of the bath is
\eq{ \label{eqGamma}
\Gamma(\w) = -\frac{1}{\pi} \tx{Im} \Delta(\w+i0^+),
}
which leads to slightly modified version of \eqref{eqNewDelta}
\eq{ \label{eqNewGamma}
\Gamma_\k(\w) = \frac{1}{\pi} \tx{Im} ( G_{\k}(\w)^{-1} + \Sigma_\k(\w)).
}

If one considers ordinary single-site DMFT, all equations remain the same and the
momentum patch index $\k$ can be dropped. In a multi-band 
calculation, the index $\k$ plays the role of the band index. For
DMFT carried out for the Bethe lattice, self-consistency can be written as 
$\Gamma(\w)=\frac{D^2}{4}A\th{imp}(\w)$,\cp{georges96}
where $A\th{imp}(\w) = -\frac{1}{\pi} \tx{Im } G\th{imp}(\w+i0^+)$. An iterative
solution is particularly simple in this case, as only the 
spectral function has to be computed 
and summations over $\bm{k}$ space are not necessary. 
In the general case, also the real part of the Green's function is needed. This can
either be accessed from the spectral function by the Kramers-Kronig relation 
or directly from the Chebyshev moments  
through\cp{weisse06}
\eq{
G\th{imp}(\w) = -\frac{i}{a} \sum_n w_n(\w') \mu_n \exp(-in\arccos(\w'))
}
where $\w'\equiv\w'(\w)$ is the rescaled frequency defined in \eqref{eqShiftScale}.
The preceding equation should be evaluated slightly away from the 
real axis $\w' \ra \w'+i0^+$.

In our computations, we parallelized the independent computations for the particle and the hole
part of the Green's (spectral) function, as well as those for different impurity sites.

\subsection{Bath discretization}

In order to represent the continuous hybridization function 
$\Delta(z)$ using a discrete chain, we use the general procedure of \ct{bulla08}
(in the notation of Ref. \onlinecite{stadler13}) adding details for the
special case of the linear discretization.

If we know the hybridization function $\Gamma(\w)$ \eqref{eqGamma} on the real axis,
the bath and coupling Hamiltonian can be written as
\eq{
H_b = \int_{-1}^{1} d\ve\, \ve a\dag_\ve a_\ve + \int_{-1}^{1} d\ve\, \sqrt{\Gamma(\ve)} (d\dag a_\ve + \tx{h.c.})
}
We discretize the Hamiltonian using a linear discretization of the bath energies
\eq{
I_n & = [\epsilon_n, \epsilon_{n+1}] ,\\
\epsilon_n & = n \Delta\epsilon + \epsilon_0 \tx{ for } n \in \{1,2,\dots,L_b\}. \non
}
For a given bath size $L_b$, 
we fix the free parameters $\epsilon_0$ and $\Delta\epsilon$ by requiring 
$\int_{\epsilon_0}^{\epsilon_{L_b}}d\omega \Gamma(\omega) 
= 0.97 \int_{-\infty}^{\infty}d\omega \Gamma(\omega)$. This
leads to outer interval borders $\epsilon_0$ and $\epsilon_{L_b}$ that
are close enough to minimize finite-size effects, and far enough apart from
each other, to contain almost the complete support of $\Gamma(\w)$.
Starting with an interval $[\epsilon_0\th{init},\epsilon_{L_b}\th{init}]$
that contains the full integrated weight of $\Gamma(\w)$, 
we repeatedly shift the boundaries by a fixed small number to shrink it down 
to the required size. In a single step, we choose the boundary, that can 
be shifted with a smaller reduction of the total integral weight. The boundary
that leads to a higher reduction is left unchanged in this step.
When using a logarithmic discretization, we defined the discretization intervals
via energies $\epsilon_m\propto \pm\Lambda^{-m}$, where $m\in[1,...,L_b/2]$.\cp{bulla08} The
specific choice of boundaries of the support is not of much importance in this case.

The discretized SIAM then couples to $L_b$ bath states created by $a_n\dag$ each 
of which corresponds to a bath energy interval $I_n$.
One approximates the continuous $H_b$ by the discrete version
\eq{
H_b & \simeq \sum_{n=1}^{L_b} \xi_n a\dag_n a_n + \sum_{n=1}^{L_b} \gamma_n (d\dag a_n + \tx{h.c.}).  \non\\
\gamma_n^2 & = \int_{I_n} d\ve\, \Gamma(\ve), \quad
\xi_n  = \frac{1}{\gamma_n^2} \int_{I_n} d\ve\, \ve \Gamma(\ve). \non
}

In order to use an MPS representation, one has to map the preceding
Hamiltonian on a chain Hamiltonian. This is done using the Lanczos
algorithm with high-precision arithmetics 
for the diagonal quadratic matrix $(\xi_n \d_{nm})_{n,m=1}^{L_b}$
applied to the initial vector $(\c_n)_{n=1}^{L_b}$. After $L_b$ 
Lanczos iterations one obtains the site potentials $\ve_i$ as the diagonal of the tridiagonal
Lanczos matrix, and the hopping terms as the side-diagonal entries $t_i$. 
The hopping term from the impurity site to the first bath chain site is the square
root of the total hybridization magnitude
$t_0^2 = \sum_n \c_n^2 = \int d\ve \Gamma(\ve)$. With these
definitions, the final chain Hamiltonian reads
\eq{
H_b \simeq \sum_{i=0}^{L_b} t_i (c\dag_{i+1} c_i + \tx{h.c.} )  +  \sum_{i=1}^{L_b} \ve_i c\dag_i c_i,
}
where the impurity site is the first site of the chain $c\dag_0 \equiv d\dag$.

An alternative method to directly obtain the bath parameters
by truncating the continued fraction expansion of the hybridization function
as put forward by \ct{karski08}, did not show any advantages but led to 
equivalent results. As the method of \ct{karski08} leads to hopping energies 
that converge to a constant far away from the impurity, while the linear 
discretization scheme leads to polynomially decreasing hopping energies,
the linear discretization method leads to a 
smaller many-body bandwidth. This allows to use smaller rescaling values in CheMPS.

\subsection{Finding the ground-state}

The first problem to solve is finding the ground state of the model Hamiltonian.

\subsubsection{Initializing the wave function}

For the two-chain layout \eqref{eqHvbdmft} of the model, 
the following problem arises: the chemical potential of both chains can be strongly 
different, in which case the particle numbers on the left $N_{\k=+}$ 
and the right $N_{\k=-}$ chain
may be strongly different. Note that the Hamiltonian of \eqref{eqHvbdmft} commutes
with $N_{\k=+}$ and $N_{\k=-}$, as the chains are merely coupled by an interaction, not 
a hopping term. If starting a DMRG groundstate search with a global random state
for such a system, convergence can be expected to be very slow, as the local 
optimization does not pick up the global potential variation. 
Even worse, the absence of an 
hopping term between the two chains prevents that during minimization the particle numbers
in the left $N_{\k=+}$ and the right $N_{\k=-}$ chain change.  
This can in principle be compensated by choosing White's \tit{mixing factor}
\cp{white05} to be large when starting to sweep, reducing it 
when being close to convergence. But still we found it impossible to 
implement a reliable automatized groundstate search under these circumstances.

The problem can be solved by using a $U=0$ solution as initial guess for the groundstate
search. One should realize that the
partition between $N_{-}$ and $N_{+}=N-N_{-}$ (where $N$ is the total particle number)
only weakly depends on the interaction
$U$:  
The total potential and hopping energies 
scale with the bath length, whereas 
the interaction energy is a single-site quantity. 
Given the system parameters $\{\ve_{\k i}\}$ and $\{t_{\k i}\}$ for each chain $\k$,
we diagonalize the $L=L_b+1$ dimensional tridiagonal single-particle representation of a single chain 
with its associated impurity site. This gives us the particle sectors $N_{\pm}$
of the groundstate of each subsystem. The $U=0$ estimate for the total
particle number sector is $N=N_{+}+N_{-}$, as in this case both subsystems are uncoupled.
Given an initial guess for the chemical potential $\mu$, one should
initialize a wave function that fulfills the $U=0$ estimates for $N$
and $N_{+}/N_{-}$. 

\subsubsection{Finding the correct symmetry sector}

As the DMFT is grand-canonical, one still needs to solve the problem
of finding the correct particle number sector for the DMRG calculation. This can be greatly
accelerated using the $U=0$ estimate for $N$, which constitutes a \tit{rigorous} upper
bound for the particle number in the interacting system. 
For a given $\mu$ one can therefore use a bisection search, 
starting with $N$, $N-\Delta N$ and $N-2\Delta N$. In case 
$N-2\Delta N$ yields the lowest energy estimate, one has to extend the search regime to lower values
of $N$. If $N$ or $N-\Delta N$ yield the lowest energy, one can continue the ordinary bisection search.
For typical interaction values, $\Delta N / N = 0.05$ is a meaningful choice. If searching
for the maximum energy state, which is necessary if one wants to determine
the full many-body bandwidth $W=E\tl{max}-E_0$, one searches for the groundstate of 
$-H$. In this case the interaction between electrons becomes attractive, and the $U=0$
solution for the particle number sector of $\ket{E\tl{max}}$  becomes a rigorous lower bound for the interacting system.

Having found the correct symmetry sector together with its groundstate for a given value of $\mu$,
one has to check whether the requirements for the local impurity densities are fulfilled
\eq{
 n - \sum_\k \expec{c_{\k}\dag c_{\k}} \stackrel{?}{=} 0. \label{eqCheckn}
}
To find the correct value of the chemical potential, a simple update of 
the chemical potential $\mu$ with the residuum of \eqref{eqCheckn} 
is usually not sufficient to achieve convergence. 
Instead, we use this method until we found a lower 
and upper bound for $\mu$ and then use a bisection again.

In some cases, the algorithm has to break its search before reaching
the required tolerance. This is when the desired chemical potential lies directly
on the boundary which separates two different particle number sectors. If this
is the case, due to the discrete nature of our model, no solution can be found.
Such a case is typically detected by observing oscillations in the 
residuum of \eqref{eqCheckn}.

When setting up the groundstate search naively, it can easily take most of the 
computation time of the calculation. 
Using the procedures just described, it usually takes only a negligible few percent of the total computation time.

\subsection{Definition of the model Hamiltonian}
\label{secHVBDMFT}

In the following, we outline the standard procedure
that eliminates the high-energy contributions in the hybridization function. 

We want to represent the non-interacting patch Green's function
\eq{
G_{0\k}(z) = \frac{1}{\abs{P_\k}} \sum_{\k\in P_\k} \frac{1}{z+\mu-\ve_{\bm{k}}},
}
by an impurity model with Green's function
$G_{0\k}\th{imp}(z) = \frac{1}{z+\mu-\Delta(z)}$,
such that
\eq{
G_{0\k}(z) = G_{0\k}\th{imp}(z).  \label{eqSelfCons0}
}
When defining the bath hybridization function naively via
\eq{
\Delta_\k(z) = z+\mu - G_{0\k}^{-1}(z),  \label{eqLambda}
}
one observes that $\Delta_\k(z) \ra \ol{t}_\k$ for $\abs{z}\ra\infty$,
when expanding for high values of $\abs{z}$, as 
\eqs{ 
G_{0\k}(z) & =  \frac{1}{z+\mu} \left( 1 + \frac{\ol{t}_\k}{z+\mu} + \mcal{O}(z^{-1}) \right), \\
G_{0\k}^{-1}(z) & = z+\mu - \ol{t}_\k + \mcal{O}(z^{-1}),   \label{eqHighEn}
}
where $\ol t_\k =  \frac{1}{\abs{P_\k}}  \sum_{k\in P_\k} \varepsilon_k$.

This means that the corresponding spectral density of the bath 
$\Gamma(\w) = -\frac{1}{\pi} \tx{Im}\,\Delta(\w+i0^+)$ has contributions at 
arbitrarily high energies and the discretization procedure that maps $\Gamma(\w)$
onto the discrete bath Hamiltonian $H_{b}$ must fail.

This problem is solved by defining an impurity model
at a shifted chemical potential $\mu \ra \mu - \ol{t}_\k$. In the 
hybridization function of this shifted impurity model
\eq{
\Delta_\k(z) = z+\mu-\ol{t}_\k - G_{0\k}^{-1}(z),  \label{eqLambda}
} 
the constant $\ol{t}_\k$ in the high-energy expansion of $G_{0\k}^{-1}(z)$ \eqref{eqHighEn} cancels out. 
It therefore approaches zero for $\abs{z}\ra\infty$ while still fulfilling \eqref{eqSelfCons0}
for $G_{0\k}\th{imp}(z) = \frac{1}{z+\mu-\ol{t}_\k-\Delta(z)}$.
As $\ol{t}_\k$ is a simple constant shift of the chemical potential,
one can as well incorporate it into the Hamiltonian description of the impurity model, 
as done in \eqref{eqHvbdmft}.

\bibliography{lit}

\begin{thebibliography}{53}%
\makeatletter
\providecommand \@ifxundefined [1]{%
 \@ifx{#1\undefined}
}%
\providecommand \@ifnum [1]{%
 \ifnum #1\expandafter \@firstoftwo
 \else \expandafter \@secondoftwo
 \fi
}%
\providecommand \@ifx [1]{%
 \ifx #1\expandafter \@firstoftwo
 \else \expandafter \@secondoftwo
 \fi
}%
\providecommand \natexlab [1]{#1}%
\providecommand \enquote  [1]{``#1''}%
\providecommand \bibnamefont  [1]{#1}%
\providecommand \bibfnamefont [1]{#1}%
\providecommand \citenamefont [1]{#1}%
\providecommand \href@noop [0]{\@secondoftwo}%
\providecommand \href [0]{\begingroup \@sanitize@url \@href}%
\providecommand \@href[1]{\@@startlink{#1}\@@href}%
\providecommand \@@href[1]{\endgroup#1\@@endlink}%
\providecommand \@sanitize@url [0]{\catcode `\\12\catcode `\$12\catcode
  `\&12\catcode `\#12\catcode `\^12\catcode `\_12\catcode `\%12\relax}%
\providecommand \@@startlink[1]{}%
\providecommand \@@endlink[0]{}%
\providecommand \url  [0]{\begingroup\@sanitize@url \@url }%
\providecommand \@url [1]{\endgroup\@href {#1}{\urlprefix }}%
\providecommand \urlprefix  [0]{URL }%
\providecommand \Eprint [0]{\href }%
\providecommand \doibase [0]{http://dx.doi.org/}%
\providecommand \selectlanguage [0]{\@gobble}%
\providecommand \bibinfo  [0]{\@secondoftwo}%
\providecommand \bibfield  [0]{\@secondoftwo}%
\providecommand \translation [1]{[#1]}%
\providecommand \BibitemOpen [0]{}%
\providecommand \bibitemStop [0]{}%
\providecommand \bibitemNoStop [0]{.\EOS\space}%
\providecommand \EOS [0]{\spacefactor3000\relax}%
\providecommand \BibitemShut  [1]{\csname bibitem#1\endcsname}%
\let\auto@bib@innerbib\@empty
\bibitem [{\citenamefont {Metzner}\ and\ \citenamefont
  {Vollhardt}(1989)}]{metzner89}%
  \BibitemOpen
  \bibfield  {author} {\bibinfo {author} {\bibfnamefont {W.}~\bibnamefont
  {Metzner}}\ and\ \bibinfo {author} {\bibfnamefont {D.}~\bibnamefont
  {Vollhardt}},\ }\href {\doibase 10.1103/PhysRevLett.62.324} {\bibfield
  {journal} {\bibinfo  {journal} {Physical Review Letters}\ }\textbf {\bibinfo
  {volume} {62}},\ \bibinfo {pages} {324} (\bibinfo {year} {1989})}\BibitemShut
  {NoStop}%
\bibitem [{\citenamefont {Georges}\ and\ \citenamefont
  {Kotliar}(1992)}]{georges92}%
  \BibitemOpen
  \bibfield  {author} {\bibinfo {author} {\bibfnamefont {A.}~\bibnamefont
  {Georges}}\ and\ \bibinfo {author} {\bibfnamefont {G.}~\bibnamefont
  {Kotliar}},\ }\href {\doibase 10.1103/PhysRevB.45.6479} {\bibfield  {journal}
  {\bibinfo  {journal} {Phys. Rev. B}\ }\textbf {\bibinfo {volume} {45}},\
  \bibinfo {pages} {6479} (\bibinfo {year} {1992})}\BibitemShut {NoStop}%
\bibitem [{\citenamefont {Georges}\ \emph {et~al.}(1996)\citenamefont
  {Georges}, \citenamefont {Kotliar}, \citenamefont {Krauth},\ and\
  \citenamefont {Rozenberg}}]{georges96}%
  \BibitemOpen
  \bibfield  {author} {\bibinfo {author} {\bibfnamefont {A.}~\bibnamefont
  {Georges}}, \bibinfo {author} {\bibfnamefont {G.}~\bibnamefont {Kotliar}},
  \bibinfo {author} {\bibfnamefont {W.}~\bibnamefont {Krauth}}, \ and\ \bibinfo
  {author} {\bibfnamefont {M.~J.}\ \bibnamefont {Rozenberg}},\ }\href {\doibase
  10.1103/RevModPhys.68.13} {\bibfield  {journal} {\bibinfo  {journal} {Rev.
  Mod. Phys.}\ }\textbf {\bibinfo {volume} {68}},\ \bibinfo {pages} {13}
  (\bibinfo {year} {1996})}\BibitemShut {NoStop}%
\bibitem [{\citenamefont {Kotliar}\ \emph {et~al.}(2006)\citenamefont
  {Kotliar}, \citenamefont {Savrasov}, \citenamefont {Haule}, \citenamefont
  {Oudovenko}, \citenamefont {Parcollet},\ and\ \citenamefont
  {Marianetti}}]{kotliar06}%
  \BibitemOpen
  \bibfield  {author} {\bibinfo {author} {\bibfnamefont {G.}~\bibnamefont
  {Kotliar}}, \bibinfo {author} {\bibfnamefont {S.}~\bibnamefont {Savrasov}},
  \bibinfo {author} {\bibfnamefont {K.}~\bibnamefont {Haule}}, \bibinfo
  {author} {\bibfnamefont {V.}~\bibnamefont {Oudovenko}}, \bibinfo {author}
  {\bibfnamefont {O.}~\bibnamefont {Parcollet}}, \ and\ \bibinfo {author}
  {\bibfnamefont {C.}~\bibnamefont {Marianetti}},\ }\href {\doibase
  10.1103/RevModPhys.78.865} {\bibfield  {journal} {\bibinfo  {journal}
  {Reviews of Modern Physics}\ }\textbf {\bibinfo {volume} {78}},\ \bibinfo
  {pages} {865} (\bibinfo {year} {2006})}\BibitemShut {NoStop}%
\bibitem [{\citenamefont {Maier}\ \emph {et~al.}(2005)\citenamefont {Maier},
  \citenamefont {Jarrell}, \citenamefont {Pruschke},\ and\ \citenamefont
  {Hettler}}]{maier05}%
  \BibitemOpen
  \bibfield  {author} {\bibinfo {author} {\bibfnamefont {T.}~\bibnamefont
  {Maier}}, \bibinfo {author} {\bibfnamefont {M.}~\bibnamefont {Jarrell}},
  \bibinfo {author} {\bibfnamefont {T.}~\bibnamefont {Pruschke}}, \ and\
  \bibinfo {author} {\bibfnamefont {M.}~\bibnamefont {Hettler}},\ }\href
  {\doibase 10.1103/RevModPhys.77.1027} {\bibfield  {journal} {\bibinfo
  {journal} {Rev. Mod. Phys.}\ }\textbf {\bibinfo {volume} {77}},\ \bibinfo
  {pages} {1027} (\bibinfo {year} {2005})}\BibitemShut {NoStop}%
\bibitem [{\citenamefont {Gull}\ \emph {et~al.}(2011)\citenamefont {Gull},
  \citenamefont {Millis}, \citenamefont {Lichtenstein}, \citenamefont
  {Rubtsov}, \citenamefont {Troyer},\ and\ \citenamefont {Werner}}]{gull11}%
  \BibitemOpen
  \bibfield  {author} {\bibinfo {author} {\bibfnamefont {E.}~\bibnamefont
  {Gull}}, \bibinfo {author} {\bibfnamefont {A.~J.}\ \bibnamefont {Millis}},
  \bibinfo {author} {\bibfnamefont {A.~I.}\ \bibnamefont {Lichtenstein}},
  \bibinfo {author} {\bibfnamefont {A.~N.}\ \bibnamefont {Rubtsov}}, \bibinfo
  {author} {\bibfnamefont {M.}~\bibnamefont {Troyer}}, \ and\ \bibinfo {author}
  {\bibfnamefont {P.}~\bibnamefont {Werner}},\ }\href {\doibase
  10.1103/RevModPhys.83.349} {\bibfield  {journal} {\bibinfo  {journal} {Rev.
  Mod. Phys.}\ }\textbf {\bibinfo {volume} {83}},\ \bibinfo {pages} {349}
  (\bibinfo {year} {2011})}\BibitemShut {NoStop}%
\bibitem [{\citenamefont {Rubtsov}\ \emph {et~al.}(2005)\citenamefont
  {Rubtsov}, \citenamefont {Savkin},\ and\ \citenamefont
  {Lichtenstein}}]{rubtsov05}%
  \BibitemOpen
  \bibfield  {author} {\bibinfo {author} {\bibfnamefont {A.}~\bibnamefont
  {Rubtsov}}, \bibinfo {author} {\bibfnamefont {V.}~\bibnamefont {Savkin}}, \
  and\ \bibinfo {author} {\bibfnamefont {A.}~\bibnamefont {Lichtenstein}},\
  }\href {\doibase 10.1103/physrevb.72.035122} {\bibfield  {journal} {\bibinfo
  {journal} {Phys. Rev. B}\ }\textbf {\bibinfo {volume} {72}},\ \bibinfo
  {pages} {035122} (\bibinfo {year} {2005})}\BibitemShut {NoStop}%
\bibitem [{\citenamefont {Gull}\ \emph {et~al.}(2008)\citenamefont {Gull},
  \citenamefont {Werner}, \citenamefont {Parcollet},\ and\ \citenamefont
  {Troyer}}]{gull08i}%
  \BibitemOpen
  \bibfield  {author} {\bibinfo {author} {\bibfnamefont {E.}~\bibnamefont
  {Gull}}, \bibinfo {author} {\bibfnamefont {P.}~\bibnamefont {Werner}},
  \bibinfo {author} {\bibfnamefont {O.}~\bibnamefont {Parcollet}}, \ and\
  \bibinfo {author} {\bibfnamefont {M.}~\bibnamefont {Troyer}},\ }\href
  {\doibase 10.1209/0295-5075/82/57003} {\bibfield  {journal} {\bibinfo
  {journal} {EPL (Europhysics Letters)}\ }\textbf {\bibinfo {volume} {82}},\
  \bibinfo {pages} {57003} (\bibinfo {year} {2008})}\BibitemShut {NoStop}%
\bibitem [{\citenamefont {Werner}\ \emph {et~al.}(2006)\citenamefont {Werner},
  \citenamefont {Comanac}, \citenamefont {de' Medici}, \citenamefont {Troyer},\
  and\ \citenamefont {Millis}}]{werner06}%
  \BibitemOpen
  \bibfield  {author} {\bibinfo {author} {\bibfnamefont {P.}~\bibnamefont
  {Werner}}, \bibinfo {author} {\bibfnamefont {A.}~\bibnamefont {Comanac}},
  \bibinfo {author} {\bibfnamefont {L.}~\bibnamefont {de' Medici}}, \bibinfo
  {author} {\bibfnamefont {M.}~\bibnamefont {Troyer}}, \ and\ \bibinfo {author}
  {\bibfnamefont {A.}~\bibnamefont {Millis}},\ }\href {\doibase
  10.1103/physrevlett.97.076405} {\bibfield  {journal} {\bibinfo  {journal}
  {Physical Review Letters}\ }\textbf {\bibinfo {volume} {97}} (\bibinfo {year}
  {2006}),\ 10.1103/physrevlett.97.076405}\BibitemShut {NoStop}%
\bibitem [{\citenamefont {Bulla}\ \emph {et~al.}(2008)\citenamefont {Bulla},
  \citenamefont {Costi},\ and\ \citenamefont {Pruschke}}]{bulla08}%
  \BibitemOpen
  \bibfield  {author} {\bibinfo {author} {\bibfnamefont {R.}~\bibnamefont
  {Bulla}}, \bibinfo {author} {\bibfnamefont {T.}~\bibnamefont {Costi}}, \ and\
  \bibinfo {author} {\bibfnamefont {T.}~\bibnamefont {Pruschke}},\ }\href
  {\doibase 10.1103/RevModPhys.80.395} {\bibfield  {journal} {\bibinfo
  {journal} {Rev. Mod. Phys.}\ }\textbf {\bibinfo {volume} {80}},\ \bibinfo
  {pages} {395} (\bibinfo {year} {2008})}\BibitemShut {NoStop}%
\bibitem [{\citenamefont {Caffarel}\ and\ \citenamefont
  {Krauth}(1994)}]{caffarel94}%
  \BibitemOpen
  \bibfield  {author} {\bibinfo {author} {\bibfnamefont {M.}~\bibnamefont
  {Caffarel}}\ and\ \bibinfo {author} {\bibfnamefont {W.}~\bibnamefont
  {Krauth}},\ }\href {\doibase 10.1103/PhysRevLett.72.1545} {\bibfield
  {journal} {\bibinfo  {journal} {Phys. Rev. Lett.}\ }\textbf {\bibinfo
  {volume} {72}},\ \bibinfo {pages} {1545} (\bibinfo {year}
  {1994})}\BibitemShut {NoStop}%
\bibitem [{\citenamefont {Granath}\ and\ \citenamefont
  {Strand}(2012)}]{granath12}%
  \BibitemOpen
  \bibfield  {author} {\bibinfo {author} {\bibfnamefont {M.}~\bibnamefont
  {Granath}}\ and\ \bibinfo {author} {\bibfnamefont {H.~U.~R.}\ \bibnamefont
  {Strand}},\ }\href {\doibase 10.1103/physrevb.86.115111} {\bibfield
  {journal} {\bibinfo  {journal} {Phys. Rev. B}\ }\textbf {\bibinfo {volume}
  {86}},\ \bibinfo {pages} {115111} (\bibinfo {year} {2012})}\BibitemShut
  {NoStop}%
\bibitem [{\citenamefont {Lu}\ \emph {et~al.}(2014)\citenamefont {Lu},
  \citenamefont {H\"oppner}, \citenamefont {Gunnarsson},\ and\ \citenamefont
  {Haverkort}}]{lu14}%
  \BibitemOpen
  \bibfield  {author} {\bibinfo {author} {\bibfnamefont {Y.}~\bibnamefont
  {Lu}}, \bibinfo {author} {\bibfnamefont {M.}~\bibnamefont {H\"oppner}},
  \bibinfo {author} {\bibfnamefont {O.}~\bibnamefont {Gunnarsson}}, \ and\
  \bibinfo {author} {\bibfnamefont {M.~W.}\ \bibnamefont {Haverkort}},\ }\href
  {http://arxiv.org/abs/1402.0807} {\bibfield  {journal} {\bibinfo  {journal}
  {ArXiv}\ ,\ \bibinfo {pages} {1402.0807}} (\bibinfo {year} {2014})},\ \Eprint
  {http://arxiv.org/abs/1402.0807} {1402.0807} \BibitemShut {NoStop}%
\bibitem [{\citenamefont {Mitchell}\ \emph {et~al.}(2014)\citenamefont
  {Mitchell}, \citenamefont {Galpin}, \citenamefont {Wilson-Fletcher},
  \citenamefont {Logan},\ and\ \citenamefont {Bulla}}]{mitchell14}%
  \BibitemOpen
  \bibfield  {author} {\bibinfo {author} {\bibfnamefont {A.~K.}\ \bibnamefont
  {Mitchell}}, \bibinfo {author} {\bibfnamefont {M.~R.}\ \bibnamefont
  {Galpin}}, \bibinfo {author} {\bibfnamefont {S.}~\bibnamefont
  {Wilson-Fletcher}}, \bibinfo {author} {\bibfnamefont {D.~E.}\ \bibnamefont
  {Logan}}, \ and\ \bibinfo {author} {\bibfnamefont {R.}~\bibnamefont
  {Bulla}},\ }\href {\doibase 10.1103/physrevb.89.121105} {\bibfield  {journal}
  {\bibinfo  {journal} {Phys. Rev. B}\ }\textbf {\bibinfo {volume} {89}},\
  \bibinfo {pages} {121105} (\bibinfo {year} {2014})}\BibitemShut {NoStop}%
\bibitem [{\citenamefont {White}(1992)}]{white92}%
  \BibitemOpen
  \bibfield  {author} {\bibinfo {author} {\bibfnamefont {S.~R.}\ \bibnamefont
  {White}},\ }\href {\doibase 10.1103/PhysRevLett.69.2863} {\bibfield
  {journal} {\bibinfo  {journal} {Phys. Rev. Lett.}\ }\textbf {\bibinfo
  {volume} {69}},\ \bibinfo {pages} {2863} (\bibinfo {year}
  {1992})}\BibitemShut {NoStop}%
\bibitem [{\citenamefont {Schollw\"ock}(2005)}]{schollwock05}%
  \BibitemOpen
  \bibfield  {author} {\bibinfo {author} {\bibfnamefont {U.}~\bibnamefont
  {Schollw\"ock}},\ }\href {\doibase 10.1103/RevModPhys.77.259} {\bibfield
  {journal} {\bibinfo  {journal} {Rev. Mod. Phys.}\ }\textbf {\bibinfo {volume}
  {77}},\ \bibinfo {pages} {259} (\bibinfo {year} {2005})}\BibitemShut
  {NoStop}%
\bibitem [{\citenamefont {Schollw\"ock}(2011)}]{schollwock11}%
  \BibitemOpen
  \bibfield  {author} {\bibinfo {author} {\bibfnamefont {U.}~\bibnamefont
  {Schollw\"ock}},\ }\href {\doibase 10.1016/j.aop.2010.09.012} {\bibfield
  {journal} {\bibinfo  {journal} {Annals of Physics}\ }\textbf {\bibinfo
  {volume} {326}},\ \bibinfo {pages} {96} (\bibinfo {year} {2011})}\BibitemShut
  {NoStop}%
\bibitem [{\citenamefont {Hallberg}(1995)}]{hallberg95}%
  \BibitemOpen
  \bibfield  {author} {\bibinfo {author} {\bibfnamefont {K.~A.}\ \bibnamefont
  {Hallberg}},\ }\href@noop {} {\bibfield  {journal} {\bibinfo  {journal}
  {Phys. Rev. B}\ }\textbf {\bibinfo {volume} {52}},\ \bibinfo {pages} {R9827}
  (\bibinfo {year} {1995})}\BibitemShut {NoStop}%
\bibitem [{\citenamefont {Garc\'ia}\ \emph {et~al.}(2004)\citenamefont
  {Garc\'ia}, \citenamefont {Hallberg},\ and\ \citenamefont
  {Rozenberg}}]{garcia04}%
  \BibitemOpen
  \bibfield  {author} {\bibinfo {author} {\bibfnamefont {D.}~\bibnamefont
  {Garc\'ia}}, \bibinfo {author} {\bibfnamefont {K.}~\bibnamefont {Hallberg}},
  \ and\ \bibinfo {author} {\bibfnamefont {M.}~\bibnamefont {Rozenberg}},\
  }\href {\doibase 10.1103/PhysRevLett.93.246403} {\bibfield  {journal}
  {\bibinfo  {journal} {Phys. Rev. Lett.}\ }\textbf {\bibinfo {volume} {93}},\
  \bibinfo {pages} {246403} (\bibinfo {year} {2004})}\BibitemShut {NoStop}%
\bibitem [{\citenamefont {Dargel}\ \emph {et~al.}(2012)\citenamefont {Dargel},
  \citenamefont {W\"ollert}, \citenamefont {Honecker}, \citenamefont
  {McCulloch}, \citenamefont {Schollw\"ock},\ and\ \citenamefont
  {Pruschke}}]{dargel12}%
  \BibitemOpen
  \bibfield  {author} {\bibinfo {author} {\bibfnamefont {P.~E.}\ \bibnamefont
  {Dargel}}, \bibinfo {author} {\bibfnamefont {A.}~\bibnamefont {W\"ollert}},
  \bibinfo {author} {\bibfnamefont {A.}~\bibnamefont {Honecker}}, \bibinfo
  {author} {\bibfnamefont {I.~P.}\ \bibnamefont {McCulloch}}, \bibinfo {author}
  {\bibfnamefont {U.}~\bibnamefont {Schollw\"ock}}, \ and\ \bibinfo {author}
  {\bibfnamefont {T.}~\bibnamefont {Pruschke}},\ }\href {\doibase
  10.1103/PhysRevB.85.205119} {\bibfield  {journal} {\bibinfo  {journal} {Phys.
  Rev. B}\ }\textbf {\bibinfo {volume} {85}},\ \bibinfo {pages} {205119}
  (\bibinfo {year} {2012})}\BibitemShut {NoStop}%
\bibitem [{\citenamefont {Wolf}(2013)}]{wolf13iii}%
  \BibitemOpen
  \bibfield  {author} {\bibinfo {author} {\bibfnamefont {F.~A.}\ \bibnamefont
  {Wolf}},\ }\href@noop {} {\bibfield  {journal} {\bibinfo  {journal}
  {unpublished}\ } (\bibinfo {year} {2013})}\BibitemShut {NoStop}%
\bibitem [{\citenamefont {K\"uhner}\ and\ \citenamefont
  {White}(1999)}]{kuhner99}%
  \BibitemOpen
  \bibfield  {author} {\bibinfo {author} {\bibfnamefont {T.}~\bibnamefont
  {K\"uhner}}\ and\ \bibinfo {author} {\bibfnamefont {S.}~\bibnamefont
  {White}},\ }\href {\doibase 10.1103/PhysRevB.60.335} {\bibfield  {journal}
  {\bibinfo  {journal} {Phys. Rev. B}\ }\textbf {\bibinfo {volume} {60}},\
  \bibinfo {pages} {335} (\bibinfo {year} {1999})}\BibitemShut {NoStop}%
\bibitem [{\citenamefont {Jeckelmann}(2002)}]{jeckelmann02}%
  \BibitemOpen
  \bibfield  {author} {\bibinfo {author} {\bibfnamefont {E.}~\bibnamefont
  {Jeckelmann}},\ }\href {\doibase 10.1103/PhysRevB.66.045114} {\bibfield
  {journal} {\bibinfo  {journal} {Physical Review B}\ }\textbf {\bibinfo
  {volume} {66}} (\bibinfo {year} {2002}),\
  10.1103/PhysRevB.66.045114}\BibitemShut {NoStop}%
\bibitem [{\citenamefont {Karski}\ \emph {et~al.}(2008)\citenamefont {Karski},
  \citenamefont {Raas},\ and\ \citenamefont {Uhrig}}]{karski08}%
  \BibitemOpen
  \bibfield  {author} {\bibinfo {author} {\bibfnamefont {M.}~\bibnamefont
  {Karski}}, \bibinfo {author} {\bibfnamefont {C.}~\bibnamefont {Raas}}, \ and\
  \bibinfo {author} {\bibfnamefont {G.~S.}\ \bibnamefont {Uhrig}},\ }\href
  {\doibase 10.1103/PhysRevB.77.075116} {\bibfield  {journal} {\bibinfo
  {journal} {Phys. Rev. B}\ }\textbf {\bibinfo {volume} {77}},\ \bibinfo
  {pages} {075116} (\bibinfo {year} {2008})}\BibitemShut {NoStop}%
\bibitem [{\citenamefont {Karski}\ \emph {et~al.}(2005)\citenamefont {Karski},
  \citenamefont {Raas},\ and\ \citenamefont {Uhrig}}]{karski05}%
  \BibitemOpen
  \bibfield  {author} {\bibinfo {author} {\bibfnamefont {M.}~\bibnamefont
  {Karski}}, \bibinfo {author} {\bibfnamefont {C.}~\bibnamefont {Raas}}, \ and\
  \bibinfo {author} {\bibfnamefont {G.~S.}\ \bibnamefont {Uhrig}},\ }\href
  {\doibase 10.1103/PhysRevB.72.113110} {\bibfield  {journal} {\bibinfo
  {journal} {Phys. Rev. B}\ }\textbf {\bibinfo {volume} {72}},\ \bibinfo
  {pages} {113110} (\bibinfo {year} {2005})}\BibitemShut {NoStop}%
\bibitem [{\citenamefont {Nishimoto}\ and\ \citenamefont
  {Jeckelmann}(2004)}]{nishimoto04}%
  \BibitemOpen
  \bibfield  {author} {\bibinfo {author} {\bibfnamefont {S.}~\bibnamefont
  {Nishimoto}}\ and\ \bibinfo {author} {\bibfnamefont {E.}~\bibnamefont
  {Jeckelmann}},\ }\href {\doibase 10.1088/0953-8984/16/4/010} {\bibfield
  {journal} {\bibinfo  {journal} {J. Phys.: Condens. Matter}\ }\textbf
  {\bibinfo {volume} {16}},\ \bibinfo {pages} {613} (\bibinfo {year}
  {2004})}\BibitemShut {NoStop}%
\bibitem [{\citenamefont {Wei{\ss}e}\ \emph {et~al.}(2006)\citenamefont
  {Wei{\ss}e}, \citenamefont {Wellein}, \citenamefont {Alvermann},\ and\
  \citenamefont {Fehske}}]{weisse06}%
  \BibitemOpen
  \bibfield  {author} {\bibinfo {author} {\bibfnamefont {A.}~\bibnamefont
  {Wei{\ss}e}}, \bibinfo {author} {\bibfnamefont {G.}~\bibnamefont {Wellein}},
  \bibinfo {author} {\bibfnamefont {A.}~\bibnamefont {Alvermann}}, \ and\
  \bibinfo {author} {\bibfnamefont {H.}~\bibnamefont {Fehske}},\ }\href
  {\doibase 10.1103/RevModPhys.78.275} {\bibfield  {journal} {\bibinfo
  {journal} {Rev. Mod. Phys.}\ }\textbf {\bibinfo {volume} {78}},\ \bibinfo
  {pages} {275} (\bibinfo {year} {2006})}\BibitemShut {NoStop}%
\bibitem [{\citenamefont {Holzner}\ \emph {et~al.}(2011)\citenamefont
  {Holzner}, \citenamefont {Weichselbaum}, \citenamefont {McCulloch},
  \citenamefont {Schollw\"ock},\ and\ \citenamefont {von Delft}}]{holzner11}%
  \BibitemOpen
  \bibfield  {author} {\bibinfo {author} {\bibfnamefont {A.}~\bibnamefont
  {Holzner}}, \bibinfo {author} {\bibfnamefont {A.}~\bibnamefont
  {Weichselbaum}}, \bibinfo {author} {\bibfnamefont {I.~P.}\ \bibnamefont
  {McCulloch}}, \bibinfo {author} {\bibfnamefont {U.}~\bibnamefont
  {Schollw\"ock}}, \ and\ \bibinfo {author} {\bibfnamefont {J.}~\bibnamefont
  {von Delft}},\ }\href {\doibase 10.1103/PhysRevB.83.195115} {\bibfield
  {journal} {\bibinfo  {journal} {Phys. Rev. B}\ }\textbf {\bibinfo {volume}
  {83}},\ \bibinfo {pages} {195115} (\bibinfo {year} {2011})}\BibitemShut
  {NoStop}%
\bibitem [{\citenamefont {Braun}\ and\ \citenamefont
  {Schmitteckert}(2013)}]{braun13}%
  \BibitemOpen
  \bibfield  {author} {\bibinfo {author} {\bibfnamefont {A.}~\bibnamefont
  {Braun}}\ and\ \bibinfo {author} {\bibfnamefont {P.}~\bibnamefont
  {Schmitteckert}},\ }\href@noop {} {\bibfield  {journal} {\bibinfo  {journal}
  {ArXiv}\ ,\ \bibinfo {pages} {1310.2724}} (\bibinfo {year} {2013})},\ \Eprint
  {http://arxiv.org/abs/1310.2724} {1310.2724} \BibitemShut {NoStop}%
\bibitem [{\citenamefont {Tiegel}\ \emph {et~al.}(2013)\citenamefont {Tiegel},
  \citenamefont {Manmana}, \citenamefont {Pruschke},\ and\ \citenamefont
  {Honecker}}]{tiegel14}%
  \BibitemOpen
  \bibfield  {author} {\bibinfo {author} {\bibfnamefont {A.~C.}\ \bibnamefont
  {Tiegel}}, \bibinfo {author} {\bibfnamefont {S.~R.}\ \bibnamefont {Manmana}},
  \bibinfo {author} {\bibfnamefont {T.}~\bibnamefont {Pruschke}}, \ and\
  \bibinfo {author} {\bibfnamefont {A.}~\bibnamefont {Honecker}},\ }\href@noop
  {} {\bibfield  {journal} {\bibinfo  {journal} {ArXiv}\ ,\ \bibinfo {pages}
  {1312.6044}} (\bibinfo {year} {2013})},\ \Eprint
  {http://arxiv.org/abs/1312.6044} {1312.6044} \BibitemShut {NoStop}%
\bibitem [{\citenamefont {Ganahl}\ \emph
  {et~al.}(2014{\natexlab{a}})\citenamefont {Ganahl}, \citenamefont
  {Thunstr\"om}, \citenamefont {Verstraete}, \citenamefont {Held},\ and\
  \citenamefont {Evertz}}]{ganahl14}%
  \BibitemOpen
  \bibfield  {author} {\bibinfo {author} {\bibfnamefont {M.}~\bibnamefont
  {Ganahl}}, \bibinfo {author} {\bibfnamefont {P.}~\bibnamefont {Thunstr\"om}},
  \bibinfo {author} {\bibfnamefont {F.}~\bibnamefont {Verstraete}}, \bibinfo
  {author} {\bibfnamefont {K.}~\bibnamefont {Held}}, \ and\ \bibinfo {author}
  {\bibfnamefont {H.~G.}\ \bibnamefont {Evertz}},\ }\href@noop {} {\bibfield
  {journal} {\bibinfo  {journal} {ArXiv}\ } (\bibinfo {year}
  {2014}{\natexlab{a}})},\ \Eprint {http://arxiv.org/abs/1403.1209} {1403.1209}
  \BibitemShut {NoStop}%
\bibitem [{\citenamefont {Daley}\ \emph {et~al.}(2004)\citenamefont {Daley},
  \citenamefont {Kollath}, \citenamefont {Schollw\"ock},\ and\ \citenamefont
  {Vidal}}]{daley04}%
  \BibitemOpen
  \bibfield  {author} {\bibinfo {author} {\bibfnamefont {A.~J.}\ \bibnamefont
  {Daley}}, \bibinfo {author} {\bibfnamefont {C.}~\bibnamefont {Kollath}},
  \bibinfo {author} {\bibfnamefont {U.}~\bibnamefont {Schollw\"ock}}, \ and\
  \bibinfo {author} {\bibfnamefont {G.}~\bibnamefont {Vidal}},\ }\href@noop {}
  {\bibfield  {journal} {\bibinfo  {journal} {J. Stat. Mech.: Theor. Exp.}\ ,\
  \bibinfo {pages} {P04005}} (\bibinfo {year} {2004})}\BibitemShut {NoStop}%
\bibitem [{\citenamefont {Vidal}(2004)}]{vidal04}%
  \BibitemOpen
  \bibfield  {author} {\bibinfo {author} {\bibfnamefont {G.}~\bibnamefont
  {Vidal}},\ }\href@noop {} {\bibfield  {journal} {\bibinfo  {journal} {Phys.
  Rev. Lett.}\ }\textbf {\bibinfo {volume} {93}},\ \bibinfo {pages} {040502}
  (\bibinfo {year} {2004})}\BibitemShut {NoStop}%
\bibitem [{\citenamefont {White}\ and\ \citenamefont {E.}(2004)}]{white04}%
  \BibitemOpen
  \bibfield  {author} {\bibinfo {author} {\bibfnamefont {S.~R.}\ \bibnamefont
  {White}}\ and\ \bibinfo {author} {\bibfnamefont {F.~A.}\ \bibnamefont {E.}},\
  }\href@noop {} {\bibfield  {journal} {\bibinfo  {journal} {Phys. Rev. Lett.}\
  }\textbf {\bibinfo {volume} {93}},\ \bibinfo {pages} {076401} (\bibinfo
  {year} {2004})}\BibitemShut {NoStop}%
\bibitem [{\citenamefont {Ganahl}\ \emph
  {et~al.}(2014{\natexlab{b}})\citenamefont {Ganahl}, \citenamefont {Aichhorn},
  \citenamefont {Thunstr\"om}, \citenamefont {Held}, \citenamefont {Evertz},\
  and\ \citenamefont {Verstraete}}]{ganahl14i}%
  \BibitemOpen
  \bibfield  {author} {\bibinfo {author} {\bibfnamefont {M.}~\bibnamefont
  {Ganahl}}, \bibinfo {author} {\bibfnamefont {M.}~\bibnamefont {Aichhorn}},
  \bibinfo {author} {\bibfnamefont {P.}~\bibnamefont {Thunstr\"om}}, \bibinfo
  {author} {\bibfnamefont {K.}~\bibnamefont {Held}}, \bibinfo {author}
  {\bibfnamefont {H.~G.}\ \bibnamefont {Evertz}}, \ and\ \bibinfo {author}
  {\bibfnamefont {F.}~\bibnamefont {Verstraete}},\ }\href@noop {} {\bibfield
  {journal} {\bibinfo  {journal} {ArXiv}\ ,\ \bibinfo {pages} {1405.6728}}
  (\bibinfo {year} {2014}{\natexlab{b}})},\ \Eprint
  {http://arxiv.org/abs/1405.6728} {1405.6728} \BibitemShut {NoStop}%
\bibitem [{\citenamefont {Lin}\ \emph {et~al.}(2013)\citenamefont {Lin},
  \citenamefont {Saad},\ and\ \citenamefont {Yang}}]{lin13}%
  \BibitemOpen
  \bibfield  {author} {\bibinfo {author} {\bibfnamefont {L.}~\bibnamefont
  {Lin}}, \bibinfo {author} {\bibfnamefont {Y.}~\bibnamefont {Saad}}, \ and\
  \bibinfo {author} {\bibfnamefont {C.}~\bibnamefont {Yang}},\ }\href
  {http://arxiv.org/abs/1308.5467v1} {\bibfield  {journal} {\bibinfo  {journal}
  {ArXiv}\ } (\bibinfo {year} {2013})},\ \Eprint
  {http://arxiv.org/abs/1308.5467} {1308.5467} \BibitemShut {NoStop}%
\bibitem [{\citenamefont {Boyd}(2001)}]{boyd01}%
  \BibitemOpen
  \bibfield  {author} {\bibinfo {author} {\bibfnamefont {J.~B.}\ \bibnamefont
  {Boyd}},\ }\href@noop {} {\emph {\bibinfo {title} {Chebyshev and Fourier
  Spectral Methods}}}\ (\bibinfo  {publisher} {Dover Publications, Mineola, New
  York},\ \bibinfo {year} {2001})\BibitemShut {NoStop}%
\bibitem [{\citenamefont {Verstraete}\ and\ \citenamefont
  {Cirac}(2006)}]{verstraete06}%
  \BibitemOpen
  \bibfield  {author} {\bibinfo {author} {\bibfnamefont {F.}~\bibnamefont
  {Verstraete}}\ and\ \bibinfo {author} {\bibfnamefont {J.}~\bibnamefont
  {Cirac}},\ }\href {\doibase 10.1103/PhysRevB.73.094423} {\bibfield  {journal}
  {\bibinfo  {journal} {Phys. Rev. B}\ }\textbf {\bibinfo {volume} {73}},\
  \bibinfo {pages} {094423} (\bibinfo {year} {2006})}\BibitemShut {NoStop}%
\bibitem [{\citenamefont {Raas}\ \emph {et~al.}(2004)\citenamefont {Raas},
  \citenamefont {Uhrig},\ and\ \citenamefont {Anders}}]{raas04}%
  \BibitemOpen
  \bibfield  {author} {\bibinfo {author} {\bibfnamefont {C.}~\bibnamefont
  {Raas}}, \bibinfo {author} {\bibfnamefont {G.~S.}\ \bibnamefont {Uhrig}}, \
  and\ \bibinfo {author} {\bibfnamefont {F.~B.}\ \bibnamefont {Anders}},\
  }\href {\doibase 10.1103/PhysRevB.69.041102} {\bibfield  {journal} {\bibinfo
  {journal} {Phys. Rev. B}\ }\textbf {\bibinfo {volume} {69}},\ \bibinfo
  {pages} {041102} (\bibinfo {year} {2004})}\BibitemShut {NoStop}%
\bibitem [{\citenamefont {Press}\ \emph {et~al.}(2007)\citenamefont {Press},
  \citenamefont {Teukolsky}, \citenamefont {Vetterling},\ and\ \citenamefont
  {Flannery}}]{numrec07}%
  \BibitemOpen
  \bibfield  {author} {\bibinfo {author} {\bibfnamefont {W.~H.}\ \bibnamefont
  {Press}}, \bibinfo {author} {\bibfnamefont {S.~A.}\ \bibnamefont
  {Teukolsky}}, \bibinfo {author} {\bibfnamefont {W.~T.}\ \bibnamefont
  {Vetterling}}, \ and\ \bibinfo {author} {\bibfnamefont {B.~P.}\ \bibnamefont
  {Flannery}},\ }\href@noop {} {\emph {\bibinfo {title} {Numerical Recipes 3rd
  Edition: The Art of Scientific Computing}}},\ \bibinfo {edition} {3rd}\ ed.\
  (\bibinfo  {publisher} {Cambridge University Press},\ \bibinfo {address} {New
  York, NY, USA},\ \bibinfo {year} {2007})\BibitemShut {NoStop}%
\bibitem [{\citenamefont {White}\ and\ \citenamefont
  {Affleck}(2008)}]{white08}%
  \BibitemOpen
  \bibfield  {author} {\bibinfo {author} {\bibfnamefont {S.}~\bibnamefont
  {White}}\ and\ \bibinfo {author} {\bibfnamefont {I.}~\bibnamefont
  {Affleck}},\ }\href {\doibase 10.1103/physrevb.77.134437} {\bibfield
  {journal} {\bibinfo  {journal} {Phys. Rev. B}\ }\textbf {\bibinfo {volume}
  {77}},\ \bibinfo {pages} {134437} (\bibinfo {year} {2008})}\BibitemShut
  {NoStop}%
\bibitem [{\citenamefont {Barthel}\ \emph {et~al.}(2009)\citenamefont
  {Barthel}, \citenamefont {Schollw\"ock},\ and\ \citenamefont
  {White}}]{barthel09}%
  \BibitemOpen
  \bibfield  {author} {\bibinfo {author} {\bibfnamefont {T.}~\bibnamefont
  {Barthel}}, \bibinfo {author} {\bibfnamefont {U.}~\bibnamefont
  {Schollw\"ock}}, \ and\ \bibinfo {author} {\bibfnamefont {S.~R.}\
  \bibnamefont {White}},\ }\href {\doibase 10.1103/PhysRevB.79.245101}
  {\bibfield  {journal} {\bibinfo  {journal} {Phys. Rev. B}\ }\textbf {\bibinfo
  {volume} {79}},\ \bibinfo {pages} {245101} (\bibinfo {year}
  {2009})}\BibitemShut {NoStop}%
\bibitem [{\citenamefont {Silver}\ and\ \citenamefont
  {R\"oder}(1997)}]{silver97}%
  \BibitemOpen
  \bibfield  {author} {\bibinfo {author} {\bibfnamefont {R.~N.}\ \bibnamefont
  {Silver}}\ and\ \bibinfo {author} {\bibfnamefont {H.}~\bibnamefont
  {R\"oder}},\ }\href {\doibase 10.1103/PhysRevE.56.4822} {\bibfield  {journal}
  {\bibinfo  {journal} {Phys. Rev. E}\ }\textbf {\bibinfo {volume} {56}},\
  \bibinfo {pages} {4822} (\bibinfo {year} {1997})}\BibitemShut {NoStop}%
\bibitem [{\citenamefont {Ferrero}\ \emph {et~al.}(2009)\citenamefont
  {Ferrero}, \citenamefont {Cornaglia}, \citenamefont {De~Leo}, \citenamefont
  {Parcollet}, \citenamefont {Kotliar},\ and\ \citenamefont
  {Georges}}]{ferrero09}%
  \BibitemOpen
  \bibfield  {author} {\bibinfo {author} {\bibfnamefont {M.}~\bibnamefont
  {Ferrero}}, \bibinfo {author} {\bibfnamefont {P.}~\bibnamefont {Cornaglia}},
  \bibinfo {author} {\bibfnamefont {L.}~\bibnamefont {De~Leo}}, \bibinfo
  {author} {\bibfnamefont {O.}~\bibnamefont {Parcollet}}, \bibinfo {author}
  {\bibfnamefont {G.}~\bibnamefont {Kotliar}}, \ and\ \bibinfo {author}
  {\bibfnamefont {A.}~\bibnamefont {Georges}},\ }\href {\doibase
  10.1103/PhysRevB.80.064501} {\bibfield  {journal} {\bibinfo  {journal}
  {Physical Review B}\ }\textbf {\bibinfo {volume} {80}},\ \bibinfo {pages}
  {064501} (\bibinfo {year} {2009})}\BibitemShut {NoStop}%
\bibitem [{\citenamefont {Ferrero}\ \emph {et~al.}(2010)\citenamefont
  {Ferrero}, \citenamefont {Parcollet}, \citenamefont {Georges}, \citenamefont
  {Kotliar},\ and\ \citenamefont {Basov}}]{ferrero10}%
  \BibitemOpen
  \bibfield  {author} {\bibinfo {author} {\bibfnamefont {M.}~\bibnamefont
  {Ferrero}}, \bibinfo {author} {\bibfnamefont {O.}~\bibnamefont {Parcollet}},
  \bibinfo {author} {\bibfnamefont {A.}~\bibnamefont {Georges}}, \bibinfo
  {author} {\bibfnamefont {G.}~\bibnamefont {Kotliar}}, \ and\ \bibinfo
  {author} {\bibfnamefont {D.~N.}\ \bibnamefont {Basov}},\ }\href {\doibase
  10.1103/PhysRevB.82.054502} {\bibfield  {journal} {\bibinfo  {journal}
  {Physical Review B}\ }\textbf {\bibinfo {volume} {82}},\ \bibinfo {pages}
  {054502} (\bibinfo {year} {2010})}\BibitemShut {NoStop}%
\bibitem [{\citenamefont {Weichselbaum}\ and\ \citenamefont {von
  Delft}(2007)}]{weichselbaum07}%
  \BibitemOpen
  \bibfield  {author} {\bibinfo {author} {\bibfnamefont {A.}~\bibnamefont
  {Weichselbaum}}\ and\ \bibinfo {author} {\bibfnamefont {J.}~\bibnamefont {von
  Delft}},\ }\href {\doibase 10.1103/PhysRevLett.99.076402} {\bibfield
  {journal} {\bibinfo  {journal} {Phys. Rev. Lett.}\ }\textbf {\bibinfo
  {volume} {99}},\ \bibinfo {pages} {076402} (\bibinfo {year}
  {2007})}\BibitemShut {NoStop}%
\bibitem [{\citenamefont {Stadler}(2013)}]{stadler13}%
  \BibitemOpen
  \bibfield  {author} {\bibinfo {author} {\bibfnamefont {K.}~\bibnamefont
  {Stadler}},\ }\emph {\bibinfo {title} {Towards exploiting non-abelian
  symmetries in the Dynamical Mean-Field Theory using the Numerical
  Renormalization Group}},\ \href
  {http://www.theorie.physik.uni-muenchen.de/lsvondelft/publications/rew_thes_book/index.html}
  {Master's thesis},\ \bibinfo  {school} {LMU Munich} (\bibinfo {year}
  {2013})\BibitemShut {NoStop}%
\bibitem [{\citenamefont {Weichselbaum}(2012)}]{weichselbaum12}%
  \BibitemOpen
  \bibfield  {author} {\bibinfo {author} {\bibfnamefont {A.}~\bibnamefont
  {Weichselbaum}},\ }\href {\doibase 10.1016/j.aop.2012.07.009} {\bibfield
  {journal} {\bibinfo  {journal} {Annals of Physics}\ }\textbf {\bibinfo
  {volume} {327}},\ \bibinfo {pages} {2972} (\bibinfo {year}
  {2012})}\BibitemShut {NoStop}%
\bibitem [{\citenamefont {Ferrero}()}]{ferrero14}%
  \BibitemOpen
  \bibfield  {author} {\bibinfo {author} {\bibfnamefont {M.}~\bibnamefont
  {Ferrero}},\ }\href@noop {} {}\bibinfo {note} {Private
  Communication}\BibitemShut {NoStop}%
\bibitem [{\citenamefont {Greger}\ \emph {et~al.}(2013)\citenamefont {Greger},
  \citenamefont {Kollar},\ and\ \citenamefont {Vollhardt}}]{greger13}%
  \BibitemOpen
  \bibfield  {author} {\bibinfo {author} {\bibfnamefont {M.}~\bibnamefont
  {Greger}}, \bibinfo {author} {\bibfnamefont {M.}~\bibnamefont {Kollar}}, \
  and\ \bibinfo {author} {\bibfnamefont {D.}~\bibnamefont {Vollhardt}},\ }\href
  {\doibase 10.1103/physrevlett.110.046403} {\bibfield  {journal} {\bibinfo
  {journal} {Physical Review Letters}\ }\textbf {\bibinfo {volume} {110}},\
  \bibinfo {pages} {046403} (\bibinfo {year} {2013})}\BibitemShut {NoStop}%
\bibitem [{\citenamefont {Raas}(2005)}]{raas05}%
  \BibitemOpen
  \bibfield  {author} {\bibinfo {author} {\bibfnamefont {C.}~\bibnamefont
  {Raas}},\ }\emph {\bibinfo {title} {Dynamic Density-Matrix Renormalization
  for the Symmetric Single Impurity Anderson Model}},\ \href
  {http://www.raas.de/thesis.html} {Ph.D. thesis},\ \bibinfo  {school}
  {University of Cologne} (\bibinfo {year} {2005})\BibitemShut {NoStop}%
\bibitem [{\citenamefont {Bulla}\ \emph {et~al.}(1998)\citenamefont {Bulla},
  \citenamefont {Hewson},\ and\ \citenamefont {Pruschke}}]{bulla98}%
  \BibitemOpen
  \bibfield  {author} {\bibinfo {author} {\bibfnamefont {R.}~\bibnamefont
  {Bulla}}, \bibinfo {author} {\bibfnamefont {A.~C.}\ \bibnamefont {Hewson}}, \
  and\ \bibinfo {author} {\bibfnamefont {T.}~\bibnamefont {Pruschke}},\ }\href
  {\doibase 10.1088/0953-8984/10/37/021} {\bibfield  {journal} {\bibinfo
  {journal} {Journal of Physics: Condensed Matter}\ }\textbf {\bibinfo {volume}
  {10}},\ \bibinfo {pages} {8365} (\bibinfo {year} {1998})}\BibitemShut
  {NoStop}%
\bibitem [{\citenamefont {White}(2005)}]{white05}%
  \BibitemOpen
  \bibfield  {author} {\bibinfo {author} {\bibfnamefont {S.~R.}\ \bibnamefont
  {White}},\ }\href {\doibase 10.1103/PhysRevB.72.180403} {\bibfield  {journal}
  {\bibinfo  {journal} {Phys. Rev. B}\ }\textbf {\bibinfo {volume} {72}},\
  \bibinfo {pages} {180403} (\bibinfo {year} {2005})}\BibitemShut {NoStop}%
\end{thebibliography}%


\end{document}